\documentclass[twocolumn]{aastex701}

\begin{document}
\title{The Lazuli Space Observatory: Opportunities for time-domain and multi-messenger astronomy}
\shorttitle{TDAMM opportunities with the Lazuli space observatory}
\shortauthors{T. Wevers et al.}

\author[0000-0002-4043-9400]{Thomas Wevers}
\affiliation{Astrophysics \& Space Center, Schmidt Sciences, New York, NY 10011, USA}
\email[]{twevers@schmidtsciences.org}  

\author[]{Thomas J. Maccarone}
\affiliation{Department of Physics \& Astronomy, Texas Tech University, Box 41051, Lubbock, TX 79409-1051, USA}
\email[]{thomas.maccarone@ttu.edu}  

\author[0000-0002-6011-0530]{Antonella Palmese}
\affiliation{McWilliams Center for Cosmology and Astrophysics, Department of Physics, Carnegie Mellon University, Pittsburgh, PA 15213, USA}
\email[]{apalmese@andrew.cmu.edu} 

\author[0000-0003-4102-380X]{David J. Sand}
\affiliation{Steward Observatory, University of Arizona, 933 North Cherry Avenue, Tucson, AZ 85721-0065, USA}
\email[]{dsand@arizona.edu} 

\author[0000-0003-3703-5154]{Suvi Gezari}
\affiliation{Department of Astronomy, University of Maryland, College Park, MD, 20742-2421, USA}
\email[]{suvi@umd.edu}

\author[0000-0002-9396-7215]{Liliana Rivera Sandoval}
\affiliation{Department of Physics and Astronomy, University of Texas Rio Grande Valley, Brownsville, TX 78520, USA}
\affiliation{South Texas Space Science Institute, University of Texas Rio Grande Valley, Brownsville, TX 78520, USA}
\email[]{liliana.riverasandoval@utrgv.edu} 

\author[]{John DiPalma}
\affiliation{Project Pearl, Schmidt Sciences, New York, NY 10011, USA}
\email[]{dipalma@projectpearl.org}
 
\author[0000-0002-0832-2974]{Griffin Hosseinzadeh}
\affiliation{Department of Astronomy \& Astrophysics, University of California, San Diego, 9500 Gilman Drive, MC 0424, La Jolla, CA
92093-0424, USA}
\email[]{ghosseinzadeh@ucsd.edu} 

\author[]{Taylor J. Hoyt}
\affiliation{Physics Division, E.O. Lawrence Berkeley National Laboratory, 1 Cyclotron Rd., Berkeley, CA, 94720, USA}
\email[]{thoyt@lbl.gov} 

\author[0000-0003-2495-8670]{Mitchell Karmen}
\affiliation{Department of Physics and Astronomy, Johns Hopkins University, 3400 N. Charles Street, Baltimore, MD 21218, USA}
\email[]{mkarmen1@jhu.edu} 

\author[0000-0002-6745-4790]{Jamie Kennea}
\affiliation{Astrophysics \& Space Center, Schmidt Sciences, New York, NY 10011, USA}
\email[]{jkennea@schmidtsciences.org} 

\author[0009-0000-4830-1484]{Keerthi Kunnumkai}
\affiliation{McWilliams Center for Cosmology and Astrophysics, Department of Physics, Carnegie Mellon University, Pittsburgh, PA 15213, USA}
\email[]{kkunnumk@andrew.cmu.edu} 

\author[0000-0002-4436-4661]{Saul Perlmutter}
\affiliation{Physics Division, E.O. Lawrence Berkeley National Laboratory, 1 Cyclotron Rd., Berkeley, CA, 94720, USA}
\affiliation{Department of Physics, University of California Berkeley, Berkeley, CA 94720, USA}
\email[]{saul@lbl.gov} 

\author[0000-0002-8121-2560]{Mickael Rigault}
\affiliation{Université Lyon 1, CNRS, IP2I Lyon, UMR 5822, Villeurbanne, France}
\email[]{m.rigault@ipnl.in2p3.fr} 

\author[0000-0002-5463-9980]{Arpita Roy}
\affiliation{Astrophysics \& Space Center, Schmidt Sciences, New York, NY 10011, USA}
\email[]{aroy@schmidtsciences.org} 

\author[0000-0001-7409-5688]{Gudmundur Stefansson}
\affiliation{Astrophysics \& Space Center, Schmidt Sciences, New York, NY 10011, USA}
\email[]{gstefansson@schmidtsciences.org} 

\author[]{Fan Yang Yang}
\affiliation{Project Pearl, Schmidt Sciences, New York, NY 10011, USA}
\email[]{fyang@projectpearl.org}

\begin{abstract}
Advancing time-domain and multi-messenger astronomy requires a multi-wavelength network of observatories capable of rapidly discovering, classifying, and characterizing transient phenomena. 
A critical gap in current capabilities is the inability to follow up faint, fast-evolving transients with sensitive, wide-band imaging and spectroscopic observations from space on timescales of minutes to hours. We discuss how the Lazuli Space Observatory will address this gap through a large collecting area, optical/NIR photometry and low-resolution integral field spectroscopy, and a rapid-response architecture with a mission requirement of $<$4 hours from trigger to first photon. Based on a latency analysis, we find a credible path to realizing response times well below this requirement, with best-case scenarios below 90 minutes under favorable conditions.
We highlight extragalactic science opportunities in currently un(der)explored parts of parameter space, including gravitational wave follow-up, kilonova characterization, supernova progenitor physics, and a wide variety of fast-evolving transients and high redshift events. We further outline new observational capabilities for Galactic time-domain science, including high frequency variability in accreting systems, precision astrometry of compact objects, and the detection of compact and ultracompact binaries, enabled by high-frequency, diffraction-limited imaging and astrometry. Together, its capabilities --- combining flagship sensitivity with response times one to two orders of magnitude faster than existing large space observatories --- position Lazuli to make transformative contributions across time-domain and multi-messenger astrophysics.
\end{abstract}



\section{Introduction} 
Astrophysics is entering an era in which the time domain -- the study of how astronomical phenomena evolve over timescales as short as seconds and as long as years -- will unlock new insights into physical processes across many subfields. From the earliest moments of explosive stellar deaths, to the accretion driven outbursts from stars and black holes, to the electromagnetic counterparts of gravitational wave sources, transient phenomena offer unique opportunities to study fundamental physics, compact objects, stellar evolution, and cosmology. 

Over the past 2 decades, large sky surveys have made the discovery of new transient events and recurring variability in astronomical objects routine. 
The bottleneck in going from photons to physics has shifted from the initial discovery phase to the ability to characterize the evolution of transient phenomena over time with sensitive and timely follow-up observations. The situation is particularly challenging for transients that evolve rapidly, and those for which observations are most informative in their earliest phases.

Each step in the process from discovery to follow-up introduces a resource cost and a latency, and significant effort is being invested in automated filtering, selection and even triggering telescope time through dedicated infrastructure (the former two steps through broker systems, and the latter step through astronomical data platforms) incorporating machine learning and artificial intelligence algorithms. 

To date, most rapid spectroscopic follow-up has relied on ground-based telescopes, which are fundamentally constrained by (among other things) the day-night cycle, atmospheric conditions, and visibility windows. These constraints introduce latency and coverage gaps that are incompatible with capturing the very early and rapid (minutes-to-hours timescale) evolution of transients consistently and for large samples.

Space-based observatories have the capability, in principle, to overcome some of these challenges, offering the stability, sensitivity, and cadence needed for time-domain science that is very challenging for ground-based observatories. Historically, however, operational constraints -- including maneuverability, thermal balance, field of regard, uplink procedures, and human-in-the-loop scheduling -- have limited the responsiveness of large space-based observatories. As a result, the most sensitive facilities such as the Hubble and James Webb space telescopes (HST/JWST) have minimum turn-around times of $\ge$ 24 hours, with rapid response available only in exceptional circumstances (typically 1--2 times per year\footnote{\url{https://hst-docs.stsci.edu/hsp/hubble-space-telescope-call-for-proposals-for-cycle-34/hst-observation-types\#HSTObservationTypes-tooproposals}}).

As time-domain surveys move towards shorter cadences (e.g. the Argus array is expected to uncover and disseminate transients on timescales of minutes, \citealt{2022PASP..134c5003L}) and larger apertures (the Rubin surveys will detect transient events fainter than ever before, \citealt{2019ApJ...873..111I}), a major gap in the existing astrophysics space fleet capability is becoming increasingly acute: no existing or planned space observatory combines the capabilities for deep imaging and optical/NIR spectroscopy, flexible scheduling, and long-term field visibility in a way that enables access to follow-up observations for faint and fast transient events at scale. In response to the growing excitement for time-domain astrophysics, next generation of space telescopes (e.g. UVEX in the UV band, \citealt{2021arXiv211115608K}) are starting to be designed to address this gap, with rapid-response capabilities in the range of 2--4 hours.

In this work we describe the opportunities enabled by the Lazuli space observatory \citep{2026arXiv260102556R}, which will provide a step-change improvement in responsiveness and flexibility for optical/NIR imaging and spectroscopy from space. Coupled with a large collecting area, this provides new and highly complementary capabilities to existing facilities at other wavelengths, enabling rapid-response science (timescales $<4$ hours) in the optical and NIR domains.  

We briefly describe the Lazuli mission and its instrument suite in Section \ref{sec:mission}. We provide current best estimates for the expected target of opportunity (ToO) workflow latencies at the Lazuli science and mission operations center and spacecraft levels (\S \ref{sec:latency}); the response time from trigger to photon collection on target is generally expected to be $<$4 hours and could be as short as 60 minutes under favorable scenarios. Section \ref{sec:science} highlights science opportunities that the Lazuli space observatory will enable; in Section \ref{sec:synergy} we look forward to the expected landscape of capabilities when Lazuli launches and the synergies we expect to enable new scientific insights. We summarize in Section \ref{sec:summary}.

\section{The Lazuli space observatory}
\label{sec:mission}
The Lazuli space observatory will deploy a 3m primary mirror in an off-axis three mirror anastigmat (TMA) design \citep{2026arXiv260102556R}. Its operational orbit will be a highly elliptical 3:1 lunar resonant orbit, and it will have an instrument suite that provides novel capabilities in time domain and multi-messenger astrophysics. This includes system requirements on relevant parameters such as response latency, command uplink and data downlink latencies, slewing speed and settling times, and scheduling flexibility. 

Observing time on Lazuli will be largely allocated through open calls to the global community, with proposals selected through a peer review process. The observatory will operate with a dynamic scheduling model that supports both standard programs and rapid-response target-of-opportunity observations, including coordinated time-domain and multi-messenger campaigns with other facilities within the Eric \& Wendy Schmidt Observatory System: the Deep Synoptic Array \citep{2019BAAS...51g.255H}, the Argus Array \citep{2022PASP..134c5003L}, and the Large Fiber Array Spectroscopic Telescope (LFAST; \citealt{2022SPIE12184E..4JB}). 

There are two instruments on board that provide capabilities relevant to time domain astronomy, including a wide-field imager and an integral field spectrograph. The full system is described in detail in \citet{2026arXiv260102556R}; here we briefly describe the relevant instrument capabilities.

\subsection{Wide-field context camera}
The wide-field context camera (WCC) provides critically sampled point spread function (PSF), diffraction-limited (at 633 nm) imaging through 23 widefield CMOS sensors, nominally equipped with white-light, Sloan-like $ugriz$ broadband, and narrowband filters including H$\alpha$, H$\beta$, He\,\textsc{ii}, O\,\textsc{iii} 5007 and N\,\textsc{ii} \citep{2026arXiv260102556R}. There will be 15 IMX455 sensors with a 2.7'$\times$1.8' field of view, as well as 8 HWK4123 low read-noise sensors with a 1.5'$\times$0.9' field of view (white light in-focus and $ugiz$ filters). Due to the TMA nature of the telescope and fixed filters for every sensor, multi-band imaging will require telescope offsets. 
The detectors support both full frame imaging (up to 3.97 Hz) as well as region of interest (ROI) imaging at frame rates up to at least 500 Hz (with the read frequency fixed per sensor), although the duty cycle (i.e. the time spent reading versus integrating) decreases at high frequency.
The expected 5$\sigma$ limiting magnitude for a 60 sec integration, calculated using the WCC exposure time calculator (Stefansson et al., in prep.) is $r=25.4$ mag.

\subsection{Integral field spectrograph}
The integral field spectrograph (IFS) instrument will provide low spectral resolution (R$\sim$100--500) over a continuous wavelength coverage from 0.4--1.7$\mu$m, sampled on one of two fields: a narrow field with a size of 2.3\arcsec$\times$4.6\arcsec\ and a platescale of 40 mas pix$^{-1}$, or a wide field  with a size of 4.4\arcsec$\times$8.6\arcsec\ with a plate scale of 80 mas pix$^{-1}$ \citep{2026arXiv260102556R}. The detector is an H4RG-10 with relatively high quantum efficiency across the broad bandpass (60\% at 800 nm) and low dark current $<$0.01 e$^-$ s$^{-1}$ pix$^{-1}$. 
The expected limiting magnitude, calculated using the {\tt slicersim} IFS simulator (Rigault et al., in prep.), to achieve a mean SNR = 5 over the entire bandpass for a 1 hour integration is $g \approx 25$ mag.

\subsection{ToO response latency and frequency}
\label{sec:latency}
In order to complement and expand upon the capabilities of existing space telescopes, Lazuli operations are being designed around end-to-end workflows that optimize system latency and flexibility at each stage of the process. Following receipt of a new ToO trigger, the science operations center (SOC) will perform automated validation and constraint checks, and generate an updated short term schedule. This new plan is passed to the mission operations center (MOC) for command generation and uplink. For the highest priority ToOs, ongoing activities (such as science observations) can be interrupted immediately to start slewing to a new target; lower priority requests can be implemented as soon as possible following completion of ongoing activities.

Based on the current design and subsystem-level analysis, the total latency can be decomposed into three main components: (i) ground-segment SOC processing and schedule regeneration ($\sim$30-60 minutes for the highest priority requests), (ii) MOC command generation and uplink ($\sim$5–10 minutes), and (iii) spacecraft slewing, settling, stabilization, and acquisition. Slewing will take up to 105 minutes for a full 180 degree (edge to edge field of regard) maneuver; settling and thermal stabilization may take up to 30 minutes but is generally expected to be on the order of minutes under nominal thermal loads. The end to end fine pointing acquisition will take up to 5 minutes (but is typically expected to be much shorter), after which photons will start being collected. These estimates yield typical response times of $\lesssim$4 hours.

Achieving these response times relies on several key design choices: (1) automation-first ToO processing with minimal human-in-the-loop intervention, (2) pre-validated observing templates to reduce validation overhead, (3) dynamic queue scheduling that can be recomputed on short timescales, and (4) near-continuous commanding enabled by the selected orbit and ground segment architecture. Under favorable conditions (slew distance $<$50 degrees with typical thermal load) and a nominal system state, best-case scenario response times approaching 60-70 minutes are expected, well below the 4 hour requirement. Additional automation at the SOC level could in principle reduce this latency even further to approach the slew/settle-limited regime.

These response times will enable very sensitive observations afforded by a large space observatory in the earliest phases of transient evolution, potentially uninterrupted and/or at very high (minutes/hours) cadences and for extended periods of time ($>$weeks) given Lazuli's 3:1 lunar resonant orbit. 

In addition to the decreased response time, the relevant aspects of Lazuli science operations are being designed with regular ToO interruptions as the baseline mode of operations. As a result, we anticipate that the number of rapid response triggers from large aperture space telescopes will increase by $\sim$1--2 orders of magnitude, from 1-2 disruptive ToOs per cycle (HST/JWST) to $\mathcal{O}(10s-100)$ such opportunities per year with Lazuli.

These capabilities will enable follow-up optical/NIR observations of rapidly evolving transients from their rising phase throughout the peak and decline phases. The entire evolution, from the first photons to late-time emission, can thus be mapped in unprecedented detail, while complementary facilities, including ground-based and space-based observatories, can be used to provide higher spectral resolution, broader wavelength coverage, and better sensitivity at late times. 

\section{Science opportunities}
\label{sec:science}
In light of these observatory capabilities, we highlight science opportunities that Lazuli uniquely enables or where it will unlock new insights, both independently and in concert with other facilities. Our goal is not to be exhaustive but rather to provide an overview of the wide variety of high impact time-domain science cases that Lazuli is expected to contribute to over its mission lifetime. 

\subsection{Kilonovae}
The coalescence of binary neutron stars (BNS) and neutron star-black hole (NSBH) binaries are the most promising gravitational wave (GW) sources currently detectable by LIGO/Virgo/KAGRA to be accompanied by an electromagnetic (EM) counterpart. Even if short $\gamma$-ray bursts (sGRBs) are emitted with all of these events, only a small fraction of these will be observable (those with on-axis jet orientations), due to relativistic beaming. This makes the approximately isotropic, optical/NIR thermal emission called a “kilonova” (KN, \citealt{metzger_kilonovae}) a promising EM counterpart to GW events. On the other hand, synchrotron emission from the jet interaction with the interstellar medium, the so-called jet afterglow, may be observable for off-axis mergers \citep{Margutti2018,2026Kaur}. 

While we have some understanding of the physics behind KNe, several aspects remain poorly constrained. In particular, our main uncertainties lie in the early blue emission \citep{arcavi18}, the NS equation of state (which governs the maximum NS mass and the expected light curves,  \citealt{Zhao_2023}), the properties of the ejecta (electron fraction, mass and velocity), the production efficiency of the different heavy $r-$process elements \citep{Kasen_2017,Watson_2019,2022Gillanders}, and the diversity of these transients \citep{kasliwal20}. 

There are two distinct channels for building kilonova samples: one that leverages the discoveries of gravitational wave detectors, and a second one that relies only on EM triggers, including short GRBs from $\gamma$-ray all-sky monitors, their afterglows, and fast-evolving orphan transients found in wide area radio/UV/optical/IR surveys. The former channel yields additional information that can break degeneracies in the EM data, but it relies on GW detectors with limited on-sky periods, horizon distances and, given the relatively low observed rate of detectable NS-NS mergers, is extremely source-starved. The latter channel provides weaker constraints for individual events but enables access to a significantly larger sample, opening a regime in which the diversity, rates and evolutionary pathways can be studied statistically. 

On the other hand, a systematic study of KNe in conjunction with GW observations can provide complementary measurements of luminosity distance, inclination angle, binary masses, and classification that are not available in EM-only studies, as it was possible for the first time with the binary neutron star merger event GW170817 \citep{Abbott_2017,Margutti_2021}. Moreover, multi-messenger observations of GW events will enable cosmological constraints of the Hubble constant through the standard siren method \citep{schutz,gw170817_nature}, which can be further improved with observations of the KN
\citep{Dhawan_2020,perez}, and even more precisely if an afterglow is observed \citep{Guidorzi_2017,Palmese:2023beh,2026ApJ..1001..157A}. 

The major contribution from Lazuli will be follow-up observations of the EM counterparts to GW events discovered by other telescopes, likely with larger Field of View (FoV; $\mathcal{O}(1-10)$ sq. deg.) instruments. At the distances that the GW detectors will reach beyond 2028, the majority of KNe can only be characterized well from large aperture space telescopes \citep{Scolnic_2017,2022ApJ...927..163C}. Moreover, faint kilonovae may be serendipitously (i.e. without a GW trigger) discovered by wide and deep surveys \citep{Scolnic_2017,Andreoni_2021,2022ApJ...927..163C}, and they can be followed up by Lazuli. 

\begin{figure}
    \centering
\includegraphics[width=0.46\textwidth]{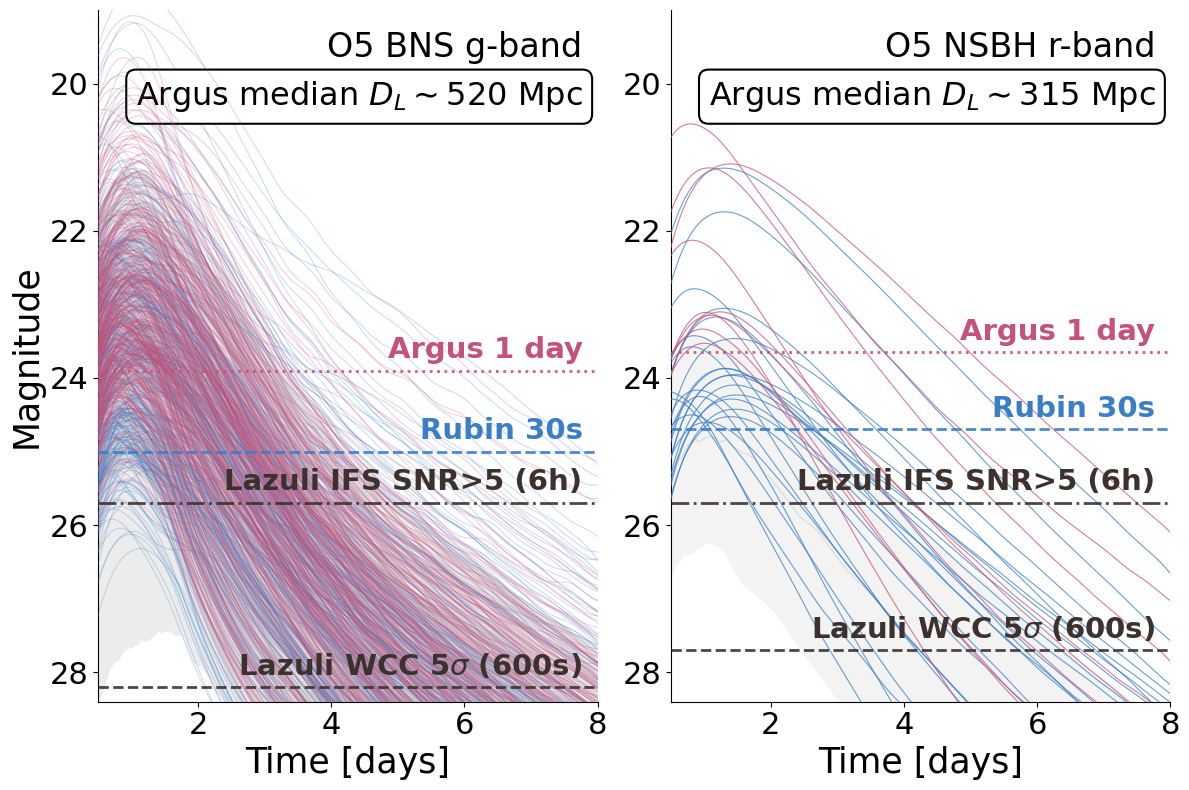}
    \caption{Kilonova lightcurves as expected from our full end-to-end simulation of GW events in O5 and their physics-motivated KN models. The left panel shows BNS KN lightcurves, while the right shows the NSBH ones, which are expected to be significantly rarer. The horizontal lines indicate, for reference, Argus, Rubin, and Lazuli depths. The Argus-detectable (assuming 1 day coadds) KNe are highlighted in pink, while the Rubin-detectable (assuming 30s exposures) ones are in blue, with $\sim 20\%$ of the former being also detectable by the latter. The KNe expected to remain undetected are shown by the shaded grey region. }
    \label{fig:fig2}
\end{figure}

Early ($<1$ day) observations are critical to discern between different mechanisms, such as radioactive decay from low-opacity dynamical ejecta, 
shock cooling powered by a jet cocoon, or cooling of wind ejecta \citep{arcavi18}. Furthermore, early KN lightcurves are sensitive to the outer ejecta structure \citep{Banerjee:2020myd,Banerjee:2023gye} and merger remnants \citep{Combi:2023yav}. The fast ToO capabilities of Lazuli will be crucial in understanding the emission mechanisms and outflows layers. Frequent ($\sim$ hours scale) $ugri$ band observations within the first day from merger may be invaluable, while also building a sample of images that can be coadded to study the host galaxy. Daily observations in $ugriz$ during day 1-10 post-merger would also help us constrain the ejecta components, considering how quickly the lightcurves are expected to fall below the typical detection limits of ground based observatories (see e.g. Fig. \ref{fig:fig2}, and Fig. 10 in \citealt{2026arXiv260102556R}).

\begin{figure}
    \centering
    \includegraphics[width=\linewidth]{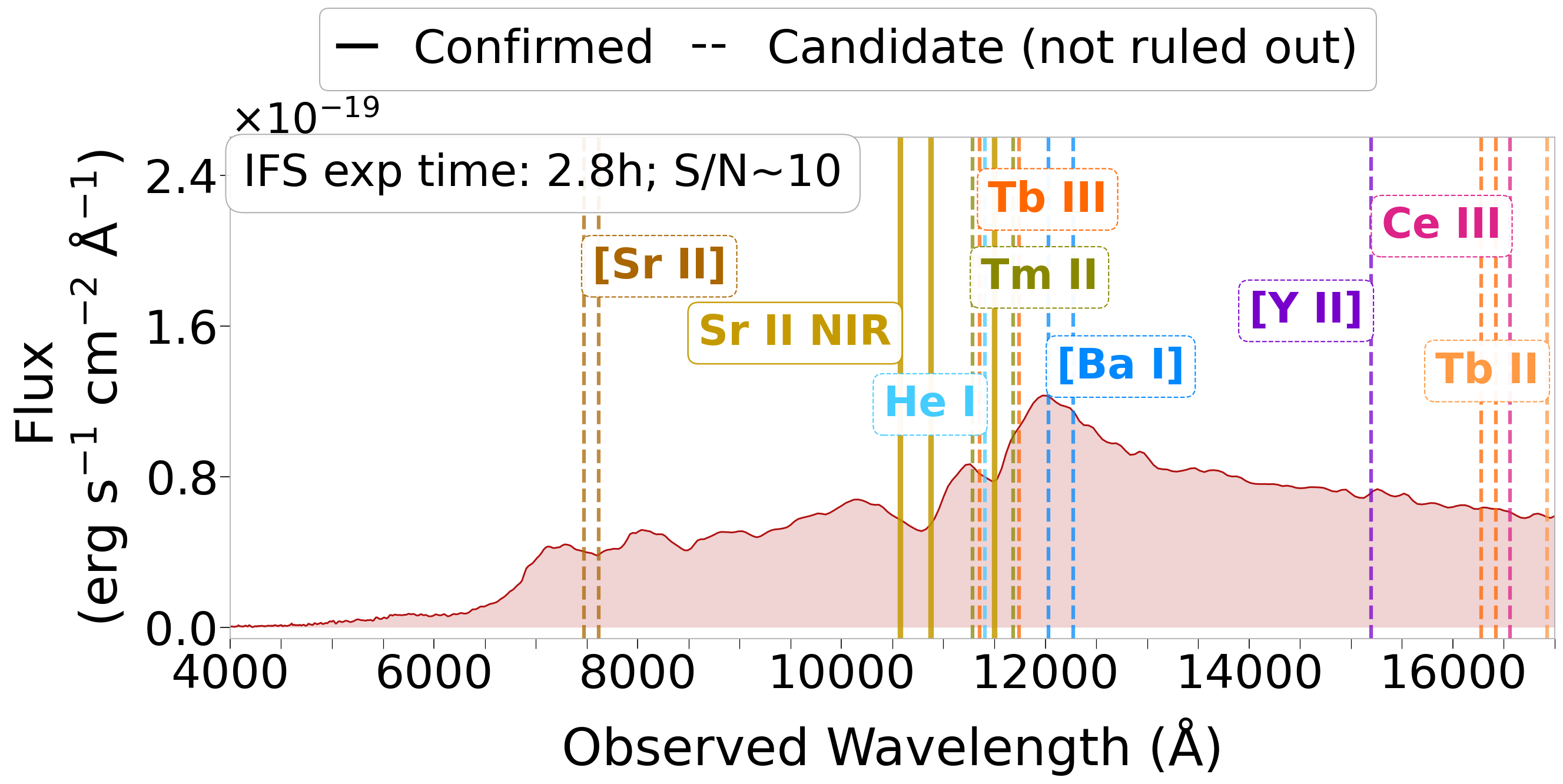}
    \caption{Simulated Lazuli/IFS BNS kilonova spectrum for a GW170817-like event at 520 Mpc, 7.4 days after merger, with an ejecta composition of Y$_{e} = 0.29$ and photospheric velocity v$_{\rm min} = 0.05c$ using the TARDIS simulations from \citet{2022Gillanders}. Solid lines indicate spectral features confirmed in the observed spectra of AT2017gfo. Dashed lines indicate candidate features consistent with the ejecta composition but whose identification remains uncertain due to incomplete near-infrared atomic data for heavy r-process elements.
 }
    \label{fig:KNspectra}
\end{figure}

Moreover, various works \citet{Watson_2019,2023A&A...675A.194S,Sneppen:2024gnt,Gillanders:2023jpd} have reported spectroscopic evidence for the presence of strontium, yttrium, and other neutron-capture elements in the AT2017gfo kilonova ejecta, supporting the interpretation that $r$-process nucleosynthesis occurred during the merger.
The identification of features from specific elements from Lazuli observations (e.g. see Fig. \ref{fig:KNspectra}) will therefore shed light on the production sites of the heaviest elements in the Universe.

Assuming a jet was launched by the merger and the viewing angle is not too off-axis, we expect the jet afterglow to peak at times later than the KN. The GW detection imposes a specific selection effect on the distribution of viewing angles for the observed GW population, typically peaking around 30 degrees with a significant population at 20-40 degrees, implying that the most likely afterglow contribution will typically peak at tens to hundreds of days post merger \citep{2026Kaur}. 

In addition to the kilonova itself, Lazuli is uniquely positioned to detect the optical emission from the afterglow, which can be used in conjunction with observations at other wavelengths to measure the synchrotron spectral index, providing important information as to the physics of the blastwave and the acceleration of electrons in the jet, as well as constraining the presence of additional dust at the merger site. 
The spectral and temporal indices provide useful ``closure relations'' \citep{2020ApJ...896..166R, 2020MNRAS.493.3521B} that are directly sensitive to the jet structure and viewing angle. This can be probed through 4-5 epochs of imaging in $r$, $i$, and/or $z$ band, log-spaced in time to capture the peak of the transient, and adjusted depending on signatures of early afterglow emission within the first 1--2 weeks after merger. Precision space-based photometry over the entire evolution will be important to discriminate between afterglow and kilonova emission.

Finally, the environment can be studied after the KN and afterglow have faded with deep $ugriz$ observations (which can also be used as templates for precision photometry of the transient), and Lazuli's IFS. This will yield invaluable information about the origin of GW sources by allowing us to probe the small scales at the location of the KN as well as faint structure in the entire host galaxy  \citep{palmese,Kilpatrick_2022,skobe2026hostgalaxies}. The blue sensitivity ($u'$-band for photometry, and $\sim0.4 ~\mu$m for spectroscopy) will be invaluable in the recovery of the star formation rate in the galaxy and at the location of the transient to study possible time delays for GW mergers (e.g. \citealt{Blanchard_2017,Zevin_2022}).

With environment data in hand, we can address the following questions: are all GW sources formed as isolated binaries (i.e., born as a pair of massive stars that each collapsed into a compact object while remaining bound), or could some be produced by dynamical interactions?  What is the delay time needed for a binary from formation to merger, and how does that compare with our expectations for massive stars in binaries undergoing stable or unstable mass transfer?

\subsubsection{Gravitational wave follow-up}
The fifth LIGO/Virgo/KAGRA observing run (O5) is expected to start in 2028 and last for about 3 years,\footnote{\url{https://emfollow.docs.ligo.org/userguide/capabilities.html}} with significant time overlap with Lazuli.
 Following a similar prescription to \citet{2025ApJ...993...15K} with the latest LIGO/Virgo/KAGRA GW rates available \citep{LIGOScientific:2025pvj} rescaled to account for the non-detection of BNS events during the fourth observing run (O4), the median number of expected BNS and NSBH detections per year is $\sim 5$ and 14, respectively.

 Accounting for the detection efficiency of two ground-based observatories (Argus and Rubin) and the theoretically predicted fraction of NSBHs expected to produce an EM counterpart ($\sim 14\%$, \citealt{Kunnumkai:2024qmw,2025ApJ...993...15K}), we estimate a detection rate of $\sim 1$ NSBH and $\sim 5$ BNS mergers with an EM counterpart per year which fall within Lazuli's follow-up capabilities (see Kunnumkai et al., in prep. for details). 
 Note that in order to identify 5 kilonovae per year, Lazuli/IFS observations may be necessary in the vetting of additional \emph{candidate} kilonovae, given the faint magnitudes of these transients (Fig. \ref{fig:fig2}). While {JWST} will possibly follow-up some of these events, disruptive ToOs are generally restricted in numbers, so that a dedicated program is needed to finally build a \emph{population} of well-characterized multi-messenger sources. A population of $\mathcal{O}(10)$ of these sources will be invaluable to understand the neutron star equation of state \citep{Nicholl_2021}, and obtain precision measurements of the Hubble constant \citep{2019PhRvL.122f1105F, Kiendrebeogo_2023,2026enap....5..557P}, which we expect to be possible by the end of the O5 run.
Delivering a systematic campaign of this scale requires the combination of rapid response, sensitivity and wavelength coverage enabled by Lazuli.

Most multimessenger events are expected to occur at 200-600 Mpc in O5, which based on simulations translates into a KN population down to $g\sim 26$ mag at 12h post merger (see Fig. \ref{fig:fig2}). For each trigger a typical dataset including the following observations should be obtained to enable the science described here:
\begin{itemize}
    \item IFS spectra:  every $\sim 4-6$ hours within the first 2 days from discovery (corresponding to $\sim 10$ minutes exposure time for a mean $S/N\sim10$ at the median distance of Argus-detectable KNe), daily after that for as long as a mean $S/N\sim5$ can be reached within a $<6$h exposure time. 
    \item $ugriz$ imaging: assuming that the EM counterpart is discovered $\lesssim12$h post-merger, acquiring broad-band imaging spaced by about 2--4h within the first day of trigger would be valuable to continuously probe the early time emission. Additional exposures spaced by $\sim 6-12$h until the kilonova is no longer detectable within up to $4\times 600$s exposures would also enable precision photometry and modeling throughout the lifetime of the transient.
    With $\sim$6 late-time epoch observations of the afterglow in the broad HWK filter for about 1 hour per epoch, it would be possible to reach the required depth to observe a 10 deg off-axis merger at $\sim 300$ Mpc (Kunnumkai et al., in prep.) and constrain the peak time, which can in turn be used to constrain the binary viewing angle.
\end{itemize}

Here we assume that the counterpart has been identified from previous observations. For reference, the Rubin Observatory's LSST alone will follow up a total of 22 BNS/NSBH events ($\sim 7$ per year) according to the latest recommendation \citep{2024arXiv241104793A}. 
ULTRASAT \citep{2024ApJ...964...74S} will also be a powerful mission to localize the EM counterpart thanks to a bright early blue component. The follow-up and characterization, especially spectroscopic, will be challenging for most ground based observatories given the expected faint magnitudes even at peak, and Lazuli can fill that gap.

\subsubsection{EM-triggered studies}
An alternative pathway to building kilonova samples is through purely EM discovery channels. In contrast to GW-triggered events, EM-selected mergers do not provide direct measurements of luminosity distance, inclination angle, or binary masses, resulting in stronger degeneracies in the interpretation of individual events. However, this channel is expected to yield significantly larger samples, probing a broader range of merger properties, environments, and viewing angles, and thus enabling population-level studies that are not accessible with GW-selected samples alone.

While GRB triggers are strongly biased toward on-axis jets due to relativistic beaming, the rapidly growing population of afterglows and orphan transients discovered independently of high-energy triggers includes a significant fraction of off-axis events. Synoptic surveys are sensitive to both on-axis and off-axis afterglows, as well as to kilonova emission (e.g.\citealt{2022ApJ...927..163C,stevenson2025strategyidentifyingverac}), which is expected to be approximately isotropic. As a result, EM-only discovery channels can access a much broader distribution of viewing angles, including systems that would be missed or poorly constrained in GW observations. Furthermore, while GRB-triggered samples preferentially select systems that successfully launch ultra-relativistic jets, optical and radio surveys are expected to uncover populations of choked jets, low-Lorentz-factor outflows, and jet-less mergers, which may not produce detectable prompt $\gamma$-ray emission.

Recent simulations of next-generation surveys in the Schmidt Observatory system \citep{2026arXiv260413650F} indicate that Argus and DSA will detect 100s of long GRB afterglows per year, as well as $\mathcal{O}(10)$ short GRB afterglows and a substantial population of orphan and fast transients, some of which will be discovered pre-peak. Lazuli can provide rapid spectroscopic confirmation for kilonova candidates among these discoveries and obtain high cadence spectroscopic and photometric observations of confirmed sources, all with the uniform spectrophotometry required to enable robust analysis across samples.

\subsection{Fast Blue Optical transients}
Fast Blue Optical Transients (FBOTs) represent a recently identified and uncommon category of extragalactic transients. They are characterized by peak bolometric luminosities above $\sim10^{43}$ erg s$^{-1}$ and rapid rise times of just a few days. The physical origin of FBOTs is still under debate, with possible explanations including engine-driven explosions, supernovae embedded in dense environments, or even tidal disruption events (TDEs).

A milestone in FBOTs studies came in 2018 with AT2018cow, the prototype of this class \citep{2018Prentice}. This event displayed several unusual properties: an exceptionally high X-ray luminosity ($\sim10^{43}$ erg s$^{-1}$) detected only three days after its optical discovery, rapid soft X-ray variability (0.3–10 keV) with timescales down to hours \citep{2018RiveraSandovalCow}, variable high-energy X-ray emission (E $> 10$ keV), an optical rise of a few days (much faster than typical supernovae), an initial blue and featureless optical spectrum, a rapid fading after peak, and short-lived (1-2 day) flare-like activity in the infrared which could be due to reprocessing of high energy emission by circumstellar material \citep[e.g.][]{2019Perley}. 

\begin{figure*}
    \centering
    \includegraphics[width=\linewidth]{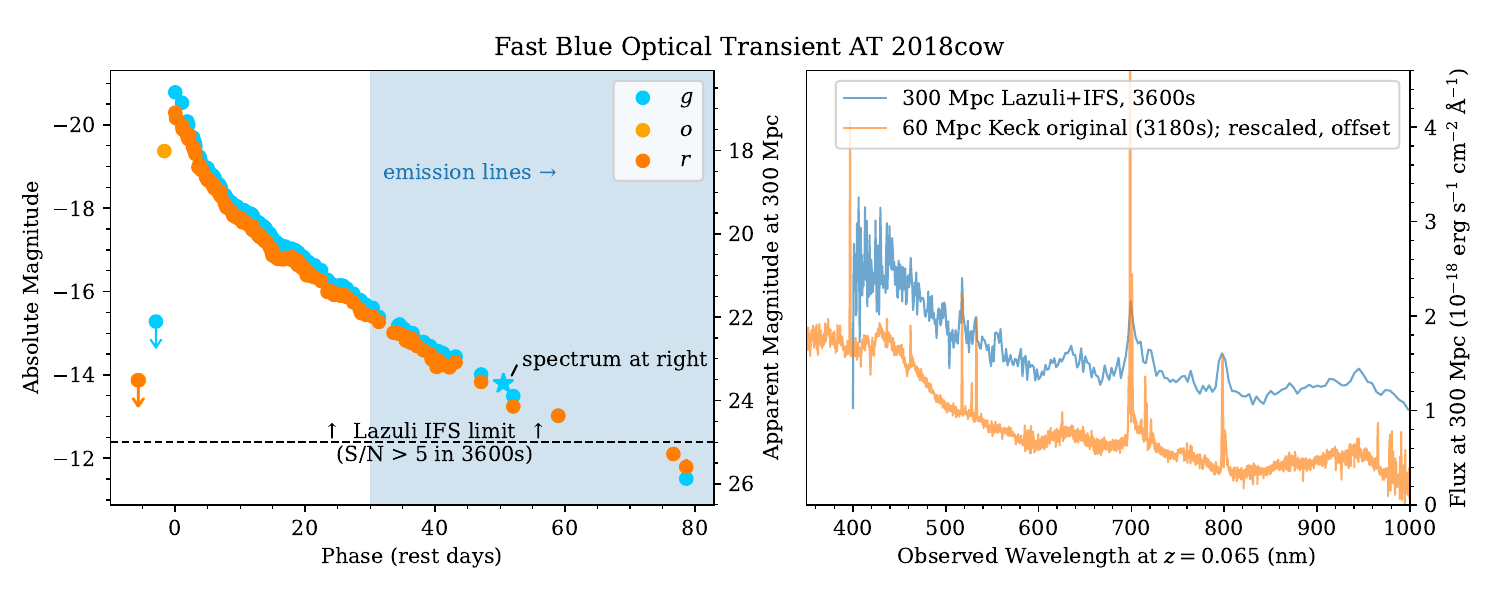}
    \caption{Light curve (\textit{left}) and spectrum (\textit{right}) of the fast blue optical transient (FBOT) AT~2018cow from \cite{2019Perley}. Emission lines emerge in the spectra of AT~2018cow around 30 days after peak, which may provide clues about the transient's power source. Lazuli's IFS will be able to obtain spectra at these phases to a $\sim$5$\times$ greater distance than AT~2018cow, potentially increasing the rate of FBOT nebular spectroscopy by $>$2 orders of magnitude.
    \label{fig:cow}}
\end{figure*}

In FBOTs, the prompt blue optical/UV emission is contemporaneous with luminous, highly variable X-ray radiation and it is followed by delayed radio and infrared emission. The optical/UV emission is generally interpreted as thermal radiation from hot ejecta, either produced directly by shock heating or by the subsequent cooling and diffusion of energy stored in the expanding ejecta \citep{2022Gottlieb}. Subsequent radio emission is consistent with synchrotron radiation produced by mildly relativistic shocks interacting with dense circumstellar material \citep{2019Ho}, and the infrared excess is often attributed to cooling ejecta or dust reprocessing of the optical/UV photons \citep{2023Metzger}. In the case of AT2018cow, where spectra were obtained out to $\sim$2 months after peak, complex hydrogen and helium emission lines at late phases may indicate ongoing interaction throughout the evolution \citep[see Fig.~\ref{fig:cow};][]{2019Perley,2019MNRAS.488.3772F}.

Since AT2018cow, only a handful of new FBOTs have been detected, and mostly at larger distances. 
Current volumetric rate estimates \citep{2023HoZTF} imply that the Rubin Observatory should discover tens to several hundred AT2018cow-like FBOTs per year, increasing the sample from fewer than ten well known examples today to hundreds over the survey lifetime.
With its high sensitivity, rapid slewing, and wide field of view, Lazuli will have the potential to follow-up and characterize from a few to several tens FBOTs per year, and will further allow investigation of the temporal and spectral correlations across different bands, which so far indicate that the observed emission is governed by complex coupled radiative and hydrodynamic processes. Photometric and spectroscopic optical/IR studies will allow to fully determine the nature of the 1-2 day duration flares as observed in AT2018cow.   Spectroscopy at late phases with the IFS will be particularly valuable, revealing to what extent interaction powers the light curve and probing properties of the circumstellar material.

On timescales that are even shorter than the $\sim$days--week typical evolution of the FBOT itself, time-resolved optical imaging has revealed bright optical flares on timescales of $\sim$minutes in the AT2018cow-like transient AT2022tsd \citep{2023Natur.623..927H} located at $z$=0.256, or 1.3 Gpc luminosity distance. These flares reach luminosities up to 10$^{44}$ erg s$^{-1}$, last for tens of minutes, have red optical colors, and do not appear to have a multi-wavelength counterpart in X-ray or radio wavelengths. The duty cycle is constrained only at the bright end of the luminosity distribution, and constraints are limited by the sensitivity of ground-based telescopes. 

The observed properties indicate that the flares are likely non-thermal in origin, with optically thin synchrotron emission as a leading candidate. Several physical mechanisms have been proposed \citep{2023Natur.623..927H}, including (relativistic) outflows or jet instabilities, episodic magnetic reconnection in a magnetar-powered engine, or variable accretion onto a compact object such as an intermediate mass black hole. A key open question is whether such flares are ubiquitous in AT2018cow-like transients, or a rare outcome of particularly energetic or favorable viewing angles. Lazuli can provide very sensitive, time-resolved observations at various phases following the initial transient, enabling constraints on the prevalence and physical diversity of this flaring behavior.

The WCC instrument on Lazuli will reach an estimated 5$\sigma$ limiting depth of $r \sim$ 25.4 mag per 60 second exposure (r$\sim$24 mag in 6 seconds), for periods up to 12 hours of uninterrupted observations. Lazuli can fully time-resolve the flare structure down to very faint magnitudes in a single photometric band. In several $\sim$hours-long exposures, Lazuli can constrain the flare luminosity distribution, energetics and duty cycle for virtually all LSST-discovered events out to redshift $z \sim 1.7$ ($z \sim 1$ for 6 second exposures), providing stringent constraints on their physical origin, and  enabling new insights into the nature of FBOTs themselves. 

At the same time, Argus can provide complementary high temporal coverage observations of all FBOTs in the Northern sky by monitoring them for long periods of time, providing constraints that will be less sensitive in depth but much better in terms of the incidence of flaring for nearby events. 
The IFS instrument will be able to provide full optical/NIR low-resolution, time-resolved spectroscopic coverage of these rapid flares on $\sim$minute timescales out to $z \sim 1$. This will provide stringent constraints on the spectral shape and its evolution, enabling a direct test of their suggested non-thermal origin. 

\subsection{Tidal disruption events / nuclear transients}
\begin{figure}
    \centering
    \includegraphics[width=\linewidth]{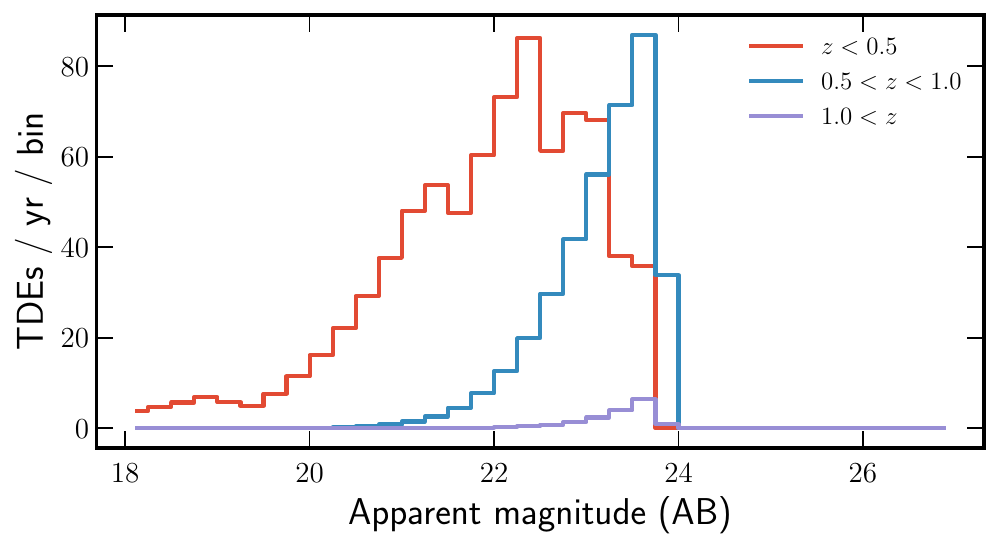}
    \caption{Apparent magnitude of tidal disruption events detected by Rubin in bins of redshift.  High-redshift TDEs ($z>1$) will be faint ($m > 23$ mag) and hostless, making Lazuli spectroscopy critical for classification.}
    \label{fig:rubin_tdes}
\end{figure}
The field of tidal disruption events (TDEs) is poised to dramatically change, with the jump in capability from discovering tens of events per year \citep{Yao2023} to thousands of events per year.  This exciting growth in TDE sample sizes will provide the possibility of probing massive black hole demographics in quiescent galaxies over a large range of masses and redshifts, and even address fundamental questions like how supermassive black holes first form in the early Universe.  However, realizing this potential requires intensive transient alert filtering, as well as systematic spectroscopic follow-up.  

While photometric classification techniques are becoming increasingly effective \citep{vanVelzen2021, Gomez2023, Stein2024}, spectroscopic measurements of redshift and emission line features are severely hampered by the lack of spectroscopic resources, and by the sensitivities required for the transients discovered by Rubin (with per epoch depths of 25th mag) and Roman (with per epoch depths of 26th mag).  This will especially be an issue for the high-redshift TDEs ($z > 1$) expected to be detected by these surveys for the first time \citep{Karmen2026}.  

Figure \ref{fig:rubin_tdes} shows the magnitude of TDEs discovered by Rubin per year in bins of redshift.  TDEs with $z > 1$ will be faint ($m > 23$ mag), and their typical galaxy hosts ($< 10^{11} M_\odot$) will be undetected in the reference images, and they will appear as hostless transients.  This makes nuclearity impossible to use as a classifier, increasing the importance of spectroscopic classification.  Lazuli has the capability to enable prompt, sensitive spectroscopic follow-up of well-vetted photometric TDE candidates for large samples of TDEs, over a broad range of redshifts (out to $z \sim 2$) and black hole masses ($10^5 M_\odot$ to $10^8 M_\odot$).

In addition to high-redshift TDEs, Rubin and Roman will be discovering fainter populations of TDEs associated with intermediate-mass black holes (IMBHs) in dwarf galaxies \citep{Ramsden2025}, and off-nuclear TDEs from massive black holes in satellite galaxies in the halo of a more massive, parent galaxy \citep{Stein2026}.  Given the empirical correlation between peak optical luminosity and black hole mass now established for the current census of optically selected TDEs \citep{Mummery2024}, we expect these lower-mass black hole TDEs to be fainter, and thus benefit from the sensitivity of Lazuli for spectroscopic follow-up.  

Furthermore, lower-mass black holes show shorter timescales \citep{Yao2023}, which imply that they will be within 0.5 mag from the peak brightness for only $\sim 1$ week after peak, again a great motivation for the prompt spectroscopic response times of Lazuli.  While IMBH TDE candidates have been detected in dwarf galaxies \citep{Angus2022, Wang2026} and off-nuclear TDEs \citep{Jin2025, Guolo2025}, the most robust signature of an IMBH would be the tidal disruption of a white dwarf (WD) \citep{Maguire2020}, which in some cases will trigger a thermonuclear transient at the time of maximum tidal compression of the WD \citep{Luminet1989, MacLeod2016}. Such sources could be photometrically identified as subluminous Type Ia SNe in the nucleus of a dwarf galaxy in the Rubin transient alert stream \citep{Gomez2023}, and promptly spectroscopically classified by Lazuli (Gezari et al., in prep.).

\begin{figure}
    \centering
    \includegraphics[width=\linewidth]{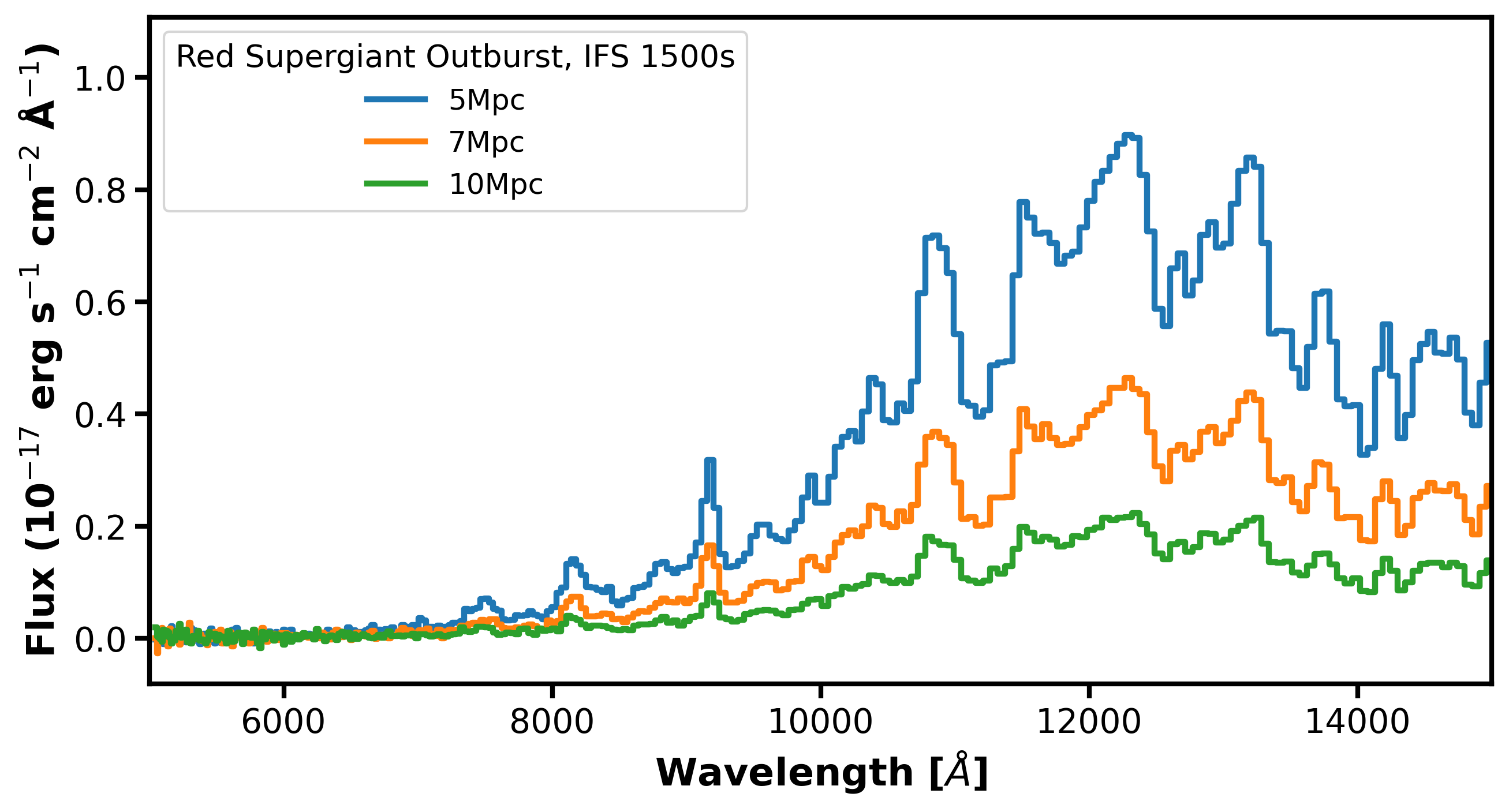}
    \caption{Simulated Lazuli/IFS data of a NIR-strong red supergiant outburst using the model of \citet{Davies22}. Such outbursts may be a pre-cursor to a core collapse supernova, depositing dense CSM around the progenitor star in the months prior to explosion. Lazuli/IFS can obtain spectroscopy of such outbursts out to $\sim$10 Mpc following detection by deep transient searches (e.g. Vera Rubin Observatory). }
    \label{fig:RSGoutburst}
\end{figure}

\subsection{Supernova science} 
\begin{figure*}
    \centering
    \includegraphics[width=0.48\linewidth]{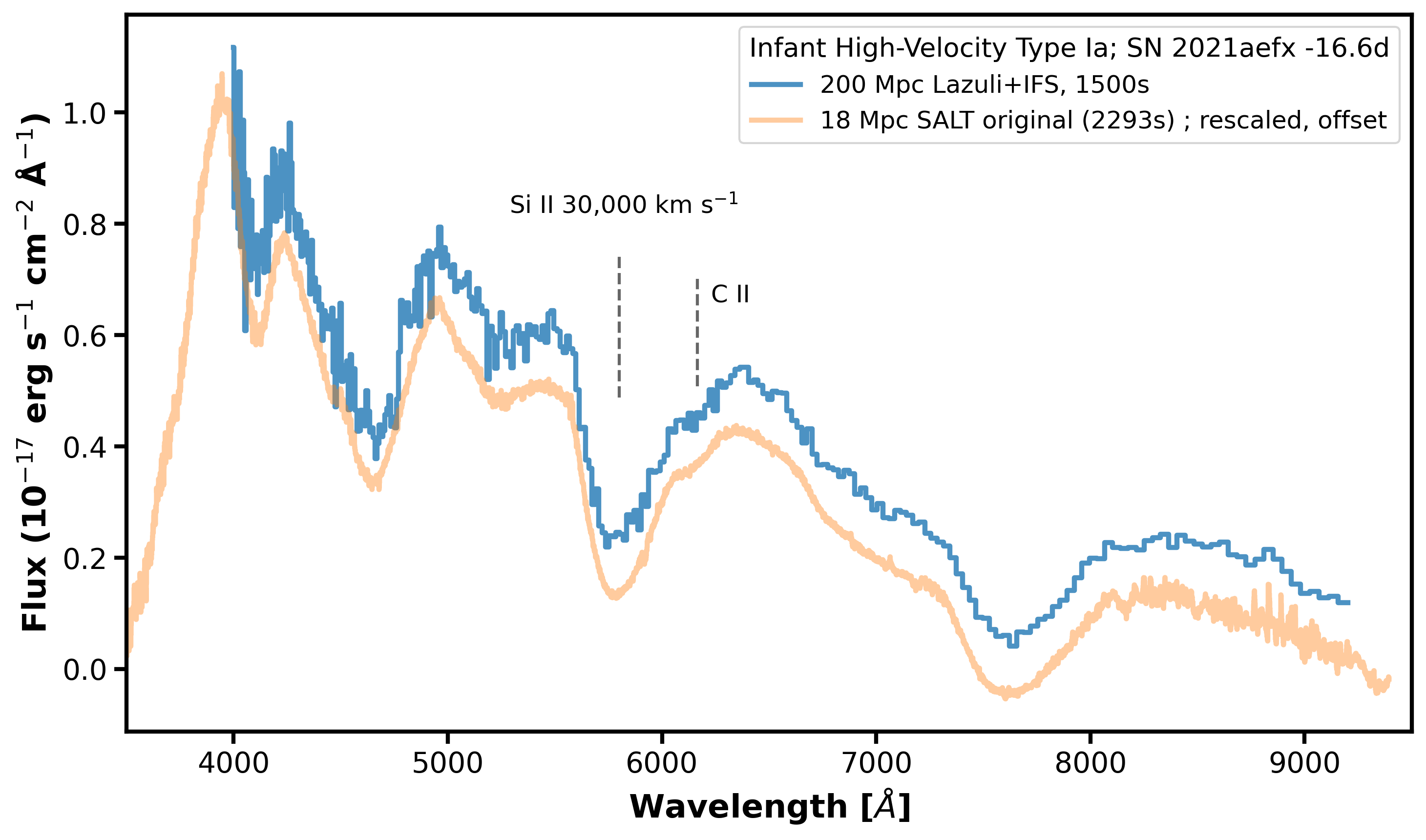}
   \includegraphics[width=0.48\linewidth]{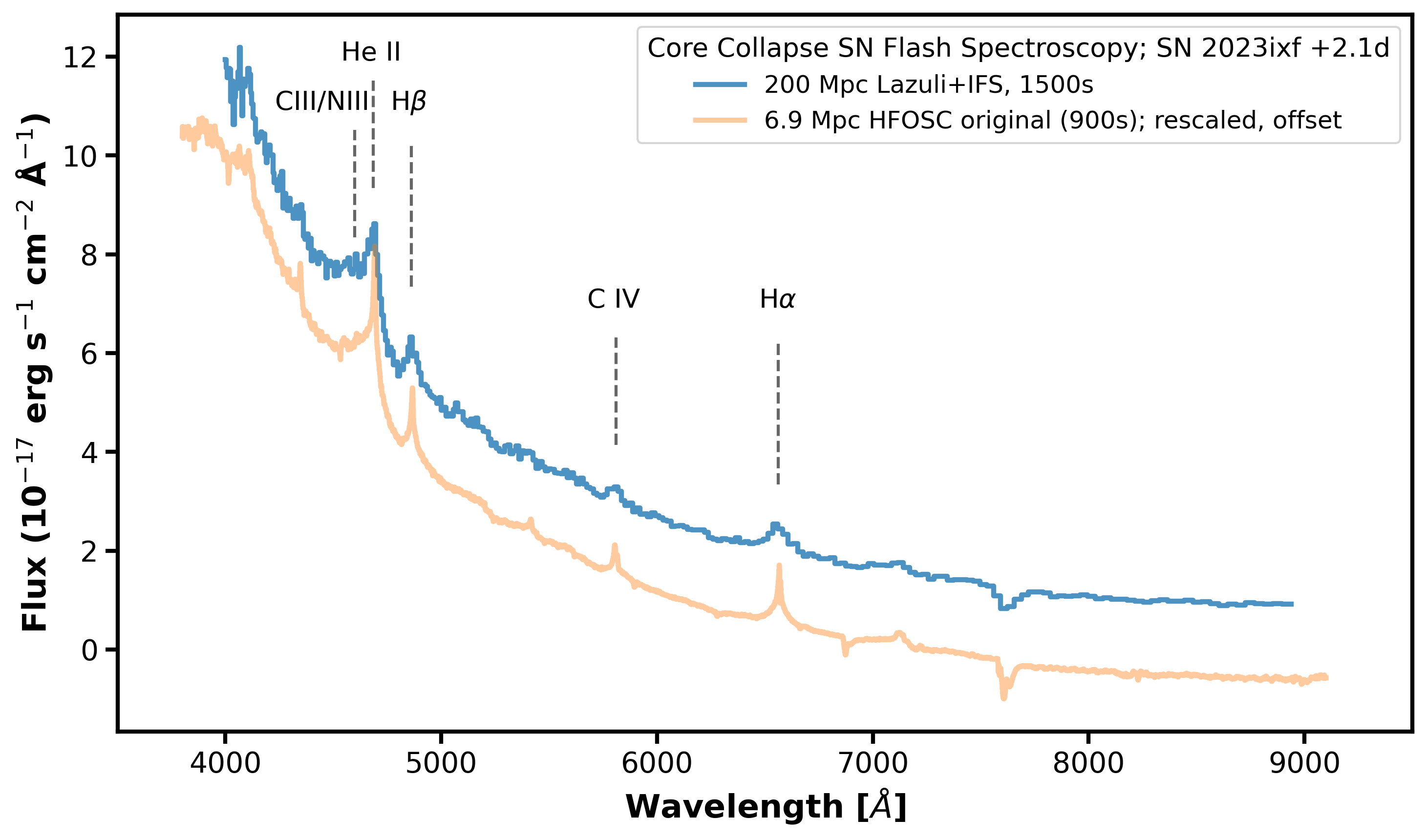}    \caption{Lazuli will enable early SN spectroscopy of moderately distant SNe in the coming time domain era, making observations that only occur once every several years at the present time more common place. Here we have taken two infant SNe spectra from the literature, and simulated Lazuli/IFS spectra at 200 Mpc with relatively short exposure times of 1500s.  {\bf Left:} An extremely early spectrum of SN2021aefx \citep[orange;][]{Hosseinzadeh22}, a type Ia SN that showed a Si\,\textsc{ii} velocity of 30,000 km s$^{-1}$, is shown alongside a simulated Lazuli spectrum.  Lazuli will be able to capture the extreme velocities now being seen in very young thermonuclear SNe, which may be due to enhanced ejecta densities \citep{Mazzali05} or circumstellar interaction \citep{Mulligan19}. {\bf Right:} A flash spectrum of SN~2023ixf \citep{Teja23}, one of the nearest core collapse SNe of the last decade.  This `flash' emission is likely a signpost of dense confined CSM around the progenitor \citep[e.g.][]{Dessart17}, and a dedicated Lazuli campaign would effectively constrain the fraction of core collapse progenitors with such a progenitor configuration.}
    \label{fig:earlySNspec}
\end{figure*}
Many questions remain about the progenitor star systems and explosion mechanisms of SNe.  One of the best ways to gain insight into SNe is to obtain observations during the first hours to days after explosion when the outer layers of the progenitor leave a spectroscopic imprint, the explosion (rather than radioactive decay) is the dominant energy source, and most circumstellar material (CSM) has not yet been overtaken by the ejecta.  

Lazuli's response time of $\lesssim$4 hours is ideally matched to the young, nearby supernovae that will be discovered when the Rubin and Argus transient streams are combined with other time domain programs.  When paired with the sensitivity of the IFS, rapid-response observations with Lazuli will probe the earliest stages of explosion in large samples of supernovae. 

\subsubsection{Core Collapse SNe}

Core collapse occurs in massive stars ($>$8 M$_{\odot}$) when fusion is unable to hold up the stellar core from its own gravity.  Core collapse supernova (CCSN) explosions are a heterogeneous group, and observers have split them up into multiple classification categories, including SN IIP, IIL, IIb, IIn and Ib/c subtypes, with many hybrid variants in between.  These various subtypes likely represent massive star end states with varying degrees of their envelope stripped at the time of explosion \citep[e.g.][]{Heger03}.  Despite growing success in determining the progenitor star systems of CCSNe via serendipitous, high-resolution imaging prior to explosion \citep{smartt15,kilpatrick16,kilpatrick21}, in the absence of such data the primary route to constraining the progenitor and environment is through pre-explosion and early-time observations of the supernova.

The final years of massive star evolution are still poorly understood. During this time period, the star may be undergoing significant mass loss and/or instabilities for various reasons \citep[e.g.][]{Chevalier12,Quataert12,SmithArnett14,Wu21} that may lead to a pre-explosion outburst or brightening.  Indeed, pre-explosion outbursts are common in type IIn SNe, occurring in up to $\sim$50\% of cases \citep[e.g.][]{Strotjohann21}. Similarly, type Ibn SNe also commonly have pre-explosion outbursts \citep[e.g.][]{Pastorello07,Dong24,Brennan24}. For normal SNe without extensive CSM, precursor outbursts are extremely rare.  For instance, only one weak outburst has been observed in a normal type IIP supernova, SN~2020tlf, starting $\sim$130 days prior to explosion and with an absolute magnitude of $\approx$$-$12 mag \citep{JG22,JG25}; several other normal type II supernovae have no outbursts to much deeper limits \citep[e.g.][]{Dong23,Shrestha24,Andrews25}.  

The lack of outbursts for normal SNII may be due to current transient survey detection limits and passbands.  Recent simulations have suggested that red supergiant (RSG) mass loss events just prior to explosion may be subject to large attenuations in the optical, with the extended atmosphere leading to molecular absorption lines and greater emission in the NIR \citep{Davies22}.  In the Rubin/LSST era, such outbursts may be detectable in redder filters in the Local Universe out to $\lesssim$10 Mpc.  In that case, the Lazuli/IFS will have a key role to play.  

In Figure~\ref{fig:RSGoutburst} we display the RSG outburst model of \citet{Davies22} at distances of D=5, 7, and 10 Mpc, using the Lazuli/IFS simulator slicersim (Rigault et al., in prep.) with an exposure time of 1800 sec.  Even at D=10 Mpc, molecular absorption features are clearly discernible in the NIR spectrum.  Although only $\sim$1 core collapse SN per year explodes within D$<$10 Mpc, triggering Lazuli on faint outbursts in the Local Universe (found via careful filtering on the LSST alert stream, for instance) may yield unprecedented data on RSG stars in the months prior to explosion.

The early time period after a CCSN explosion also offers an opportunity to learn about the progenitor star and its environment.  In the first hours to days after explosion, a significant fraction of normal CCSNe display narrow emission lines of highly ionized species before standard type II SNe features (like hydrogen P-Cygni profiles) take over in the days that follow   \citep{Khazov16,Bruch21,Bruch23}. This `flash' emission is likely due to shock breakout ionization or ejecta interaction with dense confined CSM \citep{Yaron17,Terreran22}.

The origin of the material that causes the flash emission is still unclear, as the relatively dense, confined CSM necessary exceeds what is expected from standard red supergiant winds \citep[e.g.][]{Davies22}, and there is no observational evidence for outbursts or eruptions of material in nearly all cases.  Despite dozens of observations of individual supernovae to date, the time evolution and duration of the emission lines is still not well constrained. 

For the best observed objects, it is clear that the flash features evolve on timescales of hours \citep[e.g.][]{Bostroem23,Shrestha24,Ransome26}.  In this case, rapid response spectroscopy of new, young SNe will be essential. In Figure~\ref{fig:earlySNspec} (right), we show an early spectrum of SN2023ixf \citep{Teja23}, the recent core collapse SN in M101 that displayed flash features for $\sim$7--8 days after explosion \citep{Bostroem23,JG23_23ixf}, and mark high ionization lines.  Lazuli/IFS will be able to capture flash features from SNe analogous to SN2023ixf out to a distance of $\sim$200 Mpc, even in relatively short exposure times of 1500 sec (Figure~\ref{fig:earlySNspec}). In addition, the fast response time of Lazuli ($<$4 hours) ensures that young CCSNe can be studied in the early critical hours after explosion when flash features are most likely to be found.  

Based on a search of the Transient Name Server (\url{https://www.wis-tns.org/}) within the last five years, we identified $\sim$200 normal core collapse SNe per year within the LSST footprint and D$<$200 Mpc.  Even if only a fraction of these are discovered within a day of explosion (identifiable by cross-matching multiple time domain survey streams), it should be possible to acquire a sample of $\sim$100 young CCSN with Lazuli/IFS, allowing for a $\sim$10\% measurement of the fraction of such events that exhibit flash features.  Such a measurement would provide clear insights into the environments of type II SN explosions. For a subset of the sample a spectroscopic time series would shed light on the evolution of these features, which can constrain the radial extent of the dense CSM around the progenitor star.

\subsubsection{Thermonuclear SNe} 
\label{subsec:infant}
Despite the central role of type Ia supernovae in cosmology \citep{Riess98,Perlmutter99}, we still do not have a deep understanding of SN Ia progenitors and explosion physics \citep[see][for a recent review]{Jha19}.  Several early time observational signatures, within hours to days of explosion, can shed light on the progenitor and explosion of these systems.  Here we focus on early spectroscopic signatures made possible by Lazuli/IFS observations of infant type Ia SNe.

A handful of recent SN Ia have been discovered so young that their earliest spectra display extreme silicon velocities of $\gtrsim$25,000-30,000 km s$^{-1}$ (SN2017cbv: \citealt{Hosseinzadeh17}, SN2021aefx: \citealt{Hosseinzadeh22}, SN2023bee: \citealt{Hosseinzadeh_23bee}), possibly probing the outermost layers of the ejecta.  The clearest example is SN 2021aefx, whose first spectrum from the SALT RSS spectrograph is shown in Figure~\ref{fig:earlySNspec} (left panel).  It displays a \ion{Si}{2} velocity of $\sim$30,000 km s$^{-1}$, along with a significant carbon absorption feature in the red wing of the \ion{Si}{2} line. These high velocities are unprecedented, as other SN Ia with very early spectroscopy do not show such high silicon velocity features \citep[like SN2011fe, the canonical normal type Ia SN;][]{Parrent12}.  The origin of the high velocities are unclear, but they may be due to enhanced densities in the outermost ejecta from blobs of clumpy material \citep{Mazzali05,Tanaka06}, interaction with a compact circumstellar shell \citep{Mulligan19}, or viewing-angle dependent effects of double detonation explosions \citep[e.g.][]{Boos24}.

Another early spectroscopic signal in thermonuclear supernovae comes from carbon.  The general consensus is that SNe Ia originate from the thermonuclear explosion of carbon-oxygen white dwarfs, and thus carbon is the only direct probe of unprocessed material from the progenitor system, as oxygen can be a product of carbon burning.   The incidence and quantity of unburned carbon is an important constraint on explosion models.  For instance, the pure deflagration W7 model of \citet{Nomoto84} leaves substantial carbon behind, while delayed detonation models have nearly complete carbon burning for normal SN Ia \citep{Kasen09}.  Both the pulsating class of delayed detonation models \citep{Hoflich95} and violent merger models \citep{Pakmor12} also have a large mass fraction of carbon.  
The class of sub-Chandrasekhar double detonation models, however, are characterized by the lack of remaining carbon \citep[e.g.][]{polin2019}, although a small amount may remain below the outer layer of iron group elements during a surface helium detonation \citep{Fink10}.

As we did with SN~2023ixf, we rescaled the early SN~2021aefx spectrum to a range of distances and used the IFS exposure time calculator to simulate expected data from a Lazuli campaign targeting young SNe in the Local Volume.  In Figure~\ref{fig:earlySNspec} we show the expected spectrum of an early SN~2021aefx-like supernova at a distance of 200 Mpc, showing that Lazuli/IFS will greatly expand the volume out to which very early-time SN science can be done.  Here we can clearly delineate the \ion{Si}{2} and carbon seen in the early time SALT spectrum at D=18 Mpc.  Again, a search of the TNS record indicates that $\sim$250 or more SN Ia explode within $\lesssim$200 Mpc per year.  This set of explosions, when properly filtered, will easily provide a sample that will allow for $\sim$100 or more very young SN Ia spectra with Lazuli, enabling $\lesssim$10\% measurements of the incidence of carbon and extremely high velocity features in the population, providing a benchmark to compare to simulations of thermonuclear SNe.
 
Finally, we note that Lazuli will rely on the larger time domain ecosystem to do early time supernova science.  Most notably, it will require transient alerts not just from the LSST Wide Fast Deep survey (with a cadence of $\sim$3 days), but from a {\it unified} stream of transients from multiple surveys, along with alert filters identifying transients associated with nearby galaxies, in order to identify the youngest transients in the nearby Universe.  

\subsubsection{``Calibrator'' SNe}
Type Ia supernovae (SNe~Ia) provide the most precise constraints on the Hubble constant $H_0$ \citep{Riess+2022, Freedman_2025}.
To calibrate the luminosities of SNe requires a sufficiently large number of them to have occurred within a volume accessible to primary distance indicators such as Cepheid variables. The small numbers of these calibrator SNe is the primary statistical bottleneck in $H_0$ measurements, and drives systematics due to sample selection \citep{2025arXiv250311769H, Martins_2025arXiv251114332M}. This is more fundamentally tied to the dependence of SN~Ia luminosities on global and local host properties \citep{Howell_2009, 2009ApJ...693L..76S, Rigault_2013, Rigault_2015, Rigault_2020}, heterogeneous sources of photometry (combining all possible sources since 1980 is required to reach the 1\% precision quoted today), as well as population differences between SNe~Ia \citep{Hamuy_2000, 2001ApJ...554L.193H, 2025A&A...702A.176W, 2025ApJ...986..231R}.

Lazuli/IFS observations of SNe Ia at distances $z\lesssim0.015$, too nearby for pure cosmology due to peculiar velocities but nearby enough for an independent stellar distance to be derived, would be inexpensive and potentially transformative for distance ladder $H_0$ measurements. Lazuli can provide homogeneously calibrated, spectrophotometric time-series that feed directly into new spectrophotometric methods of SN~Ia standardization that remove most, if not all, of the infamous host property dependencies \citep{Boone_2021a, Boone_2021b, Stein_2022, Ganot_2025}. The time-series could also be easily synthesized into accurate broad-band magnitudes for studies employing classical light curve approaches.

This science case would be complementary with those focused on understanding the physics of SNe~Ia, particularly via time-series of objects identified at early times (see \autoref{subsec:infant}). The objects from a hypothetical nearby SN sample classified as ``normal'' would be perfect for $H_0$ purposes. Additional novel contributions by Lazuli in this regime include the extension of the spectrophotometric model's training data (currently covering only the restframe optical) to the rest-frame NIR, providing crucial information that should, for example, be able to disentangle ``intrinsic'' and ``extrinsic'' colors (\citealt{2022MNRAS.510.3939M}; although that distinction remains ambiguous in the context of, e.g., circumstellar material).

\subsubsection{Strongly lensed SNe}
\begin{figure*}
    \centering
    \includegraphics[width=.75\linewidth]{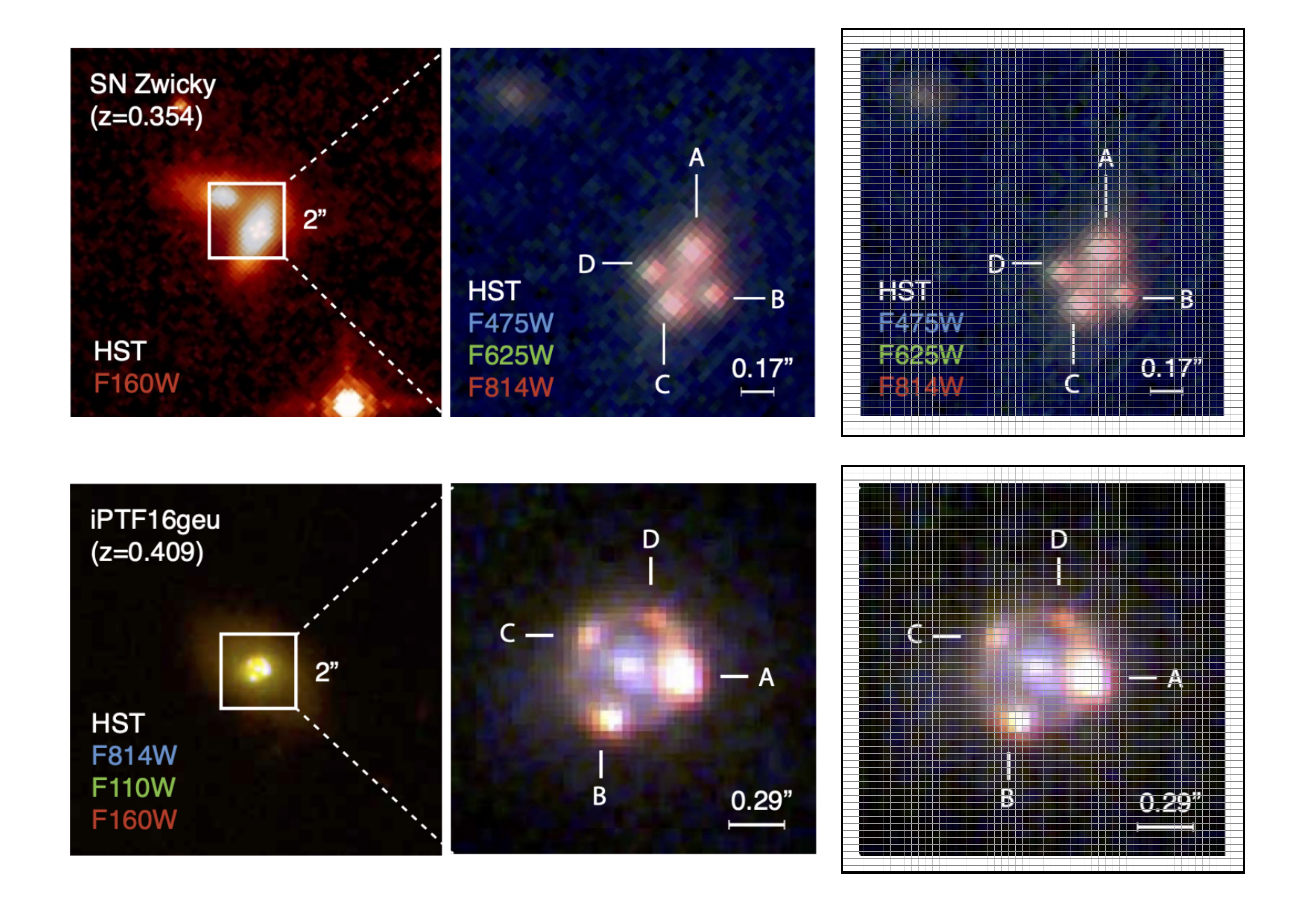}
    \caption{HST imaging of two spectrally typed SNe~Ia strongly lensed by galaxies, SN Zwicky (top) and iPTF16geu (bottom). The left panels show the wide-field images; the middle panels show a zoom into the lensed images (white box); and the right panels overlay the Lazuli IFS narrow-field grid to illustrate the expected spatial sampling. Figure adapted from \cite{goobar2025}.}
    \label{fig:slsneia}
\end{figure*}

Strong gravitational lensing of a SN by a foreground galaxy produces multiple images that arrive at different times, as each follows a distinct path through space. The time delay between images, combined with a model of the lens mass distribution, yields a measurement of the time-delay distance and therefore the Hubble constant $H_0$ \citep{Refsdal1964, Suyu+2010, Treu+2022} — completely independent of other methods and thus directly relevant to the Hubble tension between the cosmic microwave background \citep{Planck+2020} and SN~Ia distance ladder measurements \citep[e.g.,][]{Riess+2022}. Extracting this time-delay distance requires spectroscopic redshifts of the lens and source, time-delay measurements, and astrometry of the multiple images. 

To date, the majority of strongly lensed supernovae have been found through space-based surveys of galaxy clusters \citep[e.g,][]{kelly2015, rodney2021,pierel2024,frye2024}, allowing a $\sim$10\% $H_0$ measurement \citep{kelly2023, pascale2025}. In addition, two sources were discovered by ground-based surveys and spectroscopically confirmed as SNIa: PTF16gue \citep{goobar2017} and SN Zwicky \citep{goobar2023}. In both cases, these are nearby, galaxy-lensed sources ($z_\mathrm{lens}\sim0.2$) with strong magnifications (+3 mag) caused by nearly perfect earth-lens-SN alignment, as illustrated in Fig.~\ref{fig:slsneia}. In addition to these two SNe~Ia, one super-luminous and one core-collapse SN have also recently been discovered \citep[SN 2025wny and SN 2025mkn][]{taubenberger2025, johansson2025, lemon2026}, illustrating the rapid increase in discovery rate of these rare events by ground-based surveys. 

Existing surveys are detecting compact multi-image systems with short ($\lesssim1$day) time-delays, which make high precision $H_0$ measurements difficult.
However, simulations suggest that new generation surveys, including Rubin/LSST and Roman, should find hundreds of lensed SNe~Ia \citep{goldstein2019, sagues2024}; most of these will be at higher redshift than known sources and nearly all with significantly longer time delays. 

The Lazuli/IFS is particularly well suited to infer H$_0$ with high precision for these systems. Its 40 mas narrow field mode resolves even the most compact systems such as SN~Zwicky or PTF16geu (Fig.~\ref{fig:slsneia}). Spatially resolved, spectrophotometric time series enable both spectroscopic time delays \citep{johansson2021, chen2024} — by directly comparing spectral feature evolution between images — and individual image magnification constraints that tighten the lensing model. The broad wavelength coverage (400–1700\,nm) supports self-consistent analyses from nearby sources all the way out to z$\sim$3, directly relevant to early- vs.\ late-universe anchoring of the H$_0$ tension. Finally, Lazuli's rapid response enables observations at early phases when spectral feature variations are largest \citep{chen2024}.

The use of strongly-lensed supernovae for cosmology with Lazuli will by described in detail in Perlmutter et al. (in prep.).

Strongly lensed SNe also provide a unique way to constrain the progenitors of distant SN that would not otherwise be discovered early \citep[e.g.,][]{Suwa2018, Suyu+2020, Suyu+2024}. Based on the first image appearance of a lensed SN system, the occurrence of the trailing SN image(s) can be predicted, and spectra can be obtained as soon as the trailing SN image is detected in deep imaging. Given the expected redshifts of lensed SNe from Rubin/LSST at $z_{\rm SN} \sim 0.4-1.0$, the rest-frame UV will be redshifted to the optical, observable with Lazuli's IFS.

\subsection{Galactic transients and variables}
So far we have focused on extragalactic transients, but many of the same physical processes---accretion, compact object formation, and relativistic outflows---can be studied within our own Galaxy. Observations of compact binary systems not only provide key insights into supernovae, gravitational-wave sources, and jet physics, but may also open new avenues for probing fundamental astrophysics such as the distribution of dark matter in the Milky Way.

\begin{figure*}
    \centering
    \includegraphics[width=0.45\linewidth]{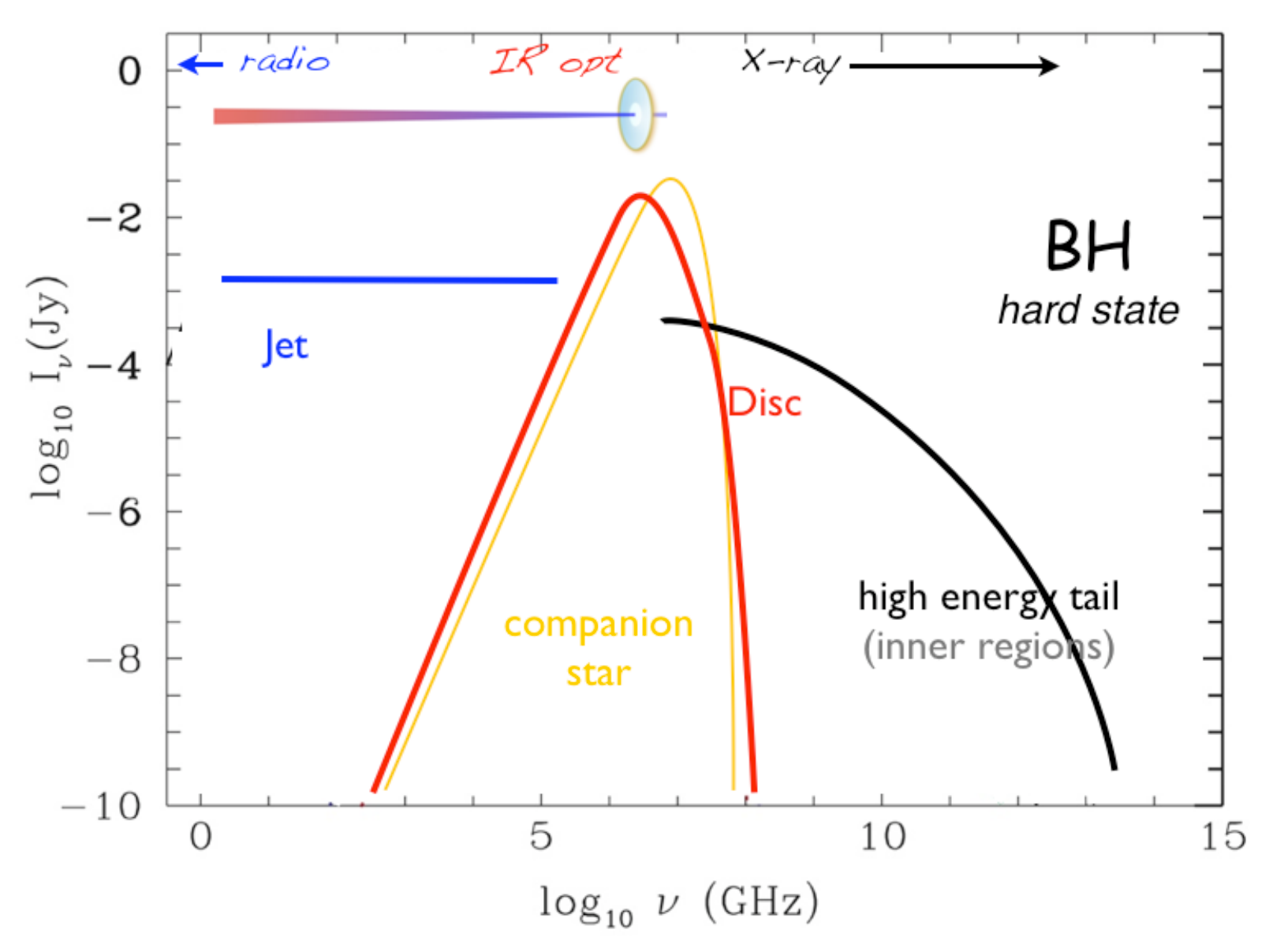}\includegraphics[width=0.35\linewidth]{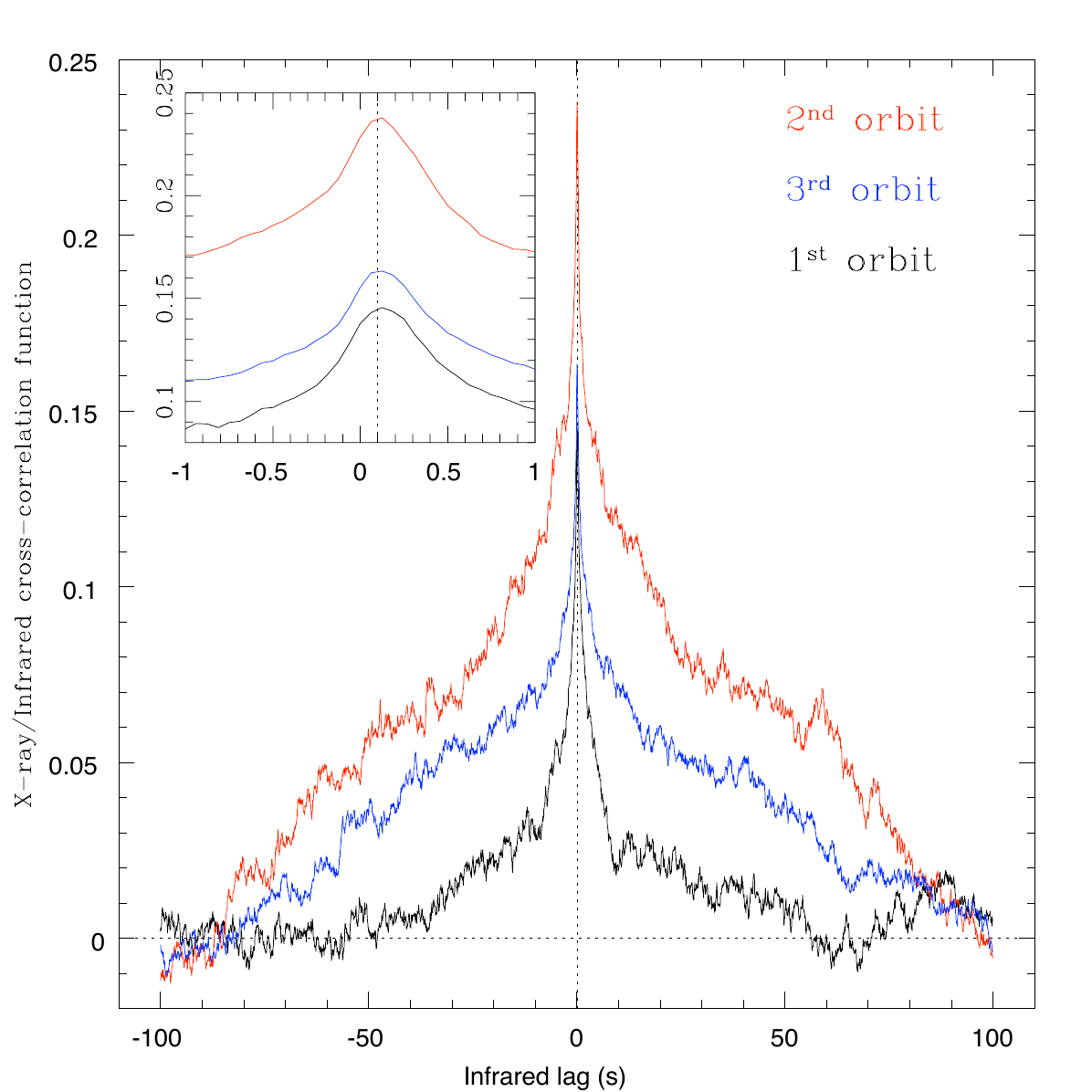}
    \caption{Left: A schematic of the spectral energy distribution of black hole X-ray binaries, showing that the jet, accretion disk, and donor star all come together in the optical-to-near-IR bands \citep{casella2015}. Right: Plots of infrared lags behind the X-rays, showing past measurements of the time lags of about 0.1 seconds of the jet emission in the X-ray behind the disk emission in X-rays (from \citep{2010MNRAS.404L..21C}).}
    \label{fig:casella}
\end{figure*}

\subsubsection{X-ray binaries: Background}

X-ray binaries are systems in which a black hole or a neutron star accretes mass from a more normal companion \citep{2023hxga.book..120B}.  The lower compact object masses in these systems lead to much faster characteristic timescales than for the supermassive black holes powering active galactic nuclei (AGN); for most AGN, the accretion disk's viscous timescale is substantially longer than humanly accessible timescales, so X-ray binaries are much better for studies as a function of global variations, rather than stochastic ones \citep{McHardy}.  A variety of discoveries have shown that large swaths of accretion physics manifest themselves in very similar ways in the two classes of objects \citep{Merloni,2003MNRAS.345L..19M,McHardy}.

In addition to accretion physics, X-ray binaries provide unique probes of the supernova explosion mechanism \citep{2022ApJ...931...94F,Burrows2025}.  Supernova light curves give excellent probes of the total mass and explosion energies of the supernovae, but only neutrinos, gravitational waves, and the properties of the compact remnants give a real understanding of what happens at the time of explosion \citep{2012ApJ...749...91F, Burrows2025}.  Developing a full understanding of the masses, spins and natal kicks of black holes, and how they are correlated, can provide probes of how supernovae explode, as can the abundances of the companion stars.

\subsubsection{X-ray binary rapid variability}
A fundamental limit on the characteristic timescales of variability that can be seen from astronomical objects comes from the differences in light travel time to an observer as a function of photon frequency. The fastest variability seen from an object can then put a lower bound on the light crossing time for the object.

In most astrophysical circumstances, variability timescales will be much slower than the light crossing times of the objects, but in cases involving propagation of material through relativistic jets, or light travel times for absorption and reprocessing of higher energy photons to produce lower energy photons, timescales of order the light crossing time of a system can be approached.  

The optical bandpass is often a place where numerous components can be contributing flux in X-ray binaries.  In faint states, the donor star typically provides of order half of the light. In bright states, the accretion disk and relativistic jet components can cross over in the optical band, and there is the added possibility of synchrotron emission from the hot, geometrically thick, optically thin component of the inflow that produces hard X-rays via Compton scattering.  Separating these components can be done through time variability -- the donor star will generally vary only on the orbital timescale, while the accretion disk will mostly vary due to reprocessing of X-rays on the light crossing time of the disk, and the jet will show emission with a lag corresponding to the light propagation time up the jet to the region where a particular wavelength can escape without synchrotron self-absorption. 

Breaking these degeneracies is important for understanding accretion in general.  By obtaining a good understanding of the relative contributions of thermal emission, synchrotron emission from the corona, and synchrotron emission from the jet, one can understand the power budgets in these components.  Understanding the thermal emission, and mapping out where it comes from using lags relative to the X-ray emission that drives it, can be used to understand the three dimensional structures of accreting binaries -- how flared they are, how big the bulges are from where the accretion stream impacts the outer disk, and how much reprocessing takes place in the donor star. 

Lazuli's sensitivity, combined with the ability to perform high frequency ($\sim$milli-seconds) imaging in multiple bands, will enable measurements of the time lag as a function of frequency even for significantly reddened (and hence faint) systems.
Generally speaking, these systems will be stationary over timescales of $\sim$ days, allowing observations at different wavelengths to be made consecutively, and for the $\sim$ second timescales of interest, data simultaneous with X-ray observations for a few tens of minutes per band will be sufficient to constrain the time lags. All objects will benefit greatly from the ability to obtain strictly simultaneous data without risk of loss due to weather.

\subsubsection{X-ray binary astrometry}
The light curves of supernovae reveal only the mass and energy of the ejecta, not the mechanism by which the energy was supplied.  The actual explosion mechanism imprints itself on the gravitational wave and neutrino bursts at the time of the supernova (which cannot be detected in the near-term except for Local Group supernovae) and on the properties of the compact remnants.  The core properties of the compact remnants to be studied are their masses, birth spins and natal kick velocities.

These properties have been notoriously difficult to measure.  Many X-ray binaries are too faint in the optical band for Gaia detections.  Masses can be estimated from emission line spectroscopy, using a series of correlations developed in the past decade \citep{2015ApJ...808...80C, 2016ApJ...822...99C, 2022MNRAS.516.2023C}.  While distances can generally be estimated by indirect means, {\it proper motions fundamentally require high angular resolution data.}  In the cases where parallaxes can be estimated by Lazuli, that will improve our knowledge of the systems.

Some of this work can be done in the radio band, but generally X-ray binaries are bright enough for precise radio detections only in outbursts, and the outbursts are not always long enough to allow a sufficient time baseline for proper motion measurements, especially in the Southern hemisphere where the sensitivity of radio telescopes is more limited, and even the upcoming Square Kilometer Array will lack the angular resolution to obtain precise proper motions in the radio band.

A variety of scenarios exist for producing stellar mass black holes, some with visible supernovae, some without.  The masses, kicks, and spins all trace the mechanisms \citep{Burrows2025}, so that mapping out the range of combinations of these parameters can indicate which of these mechanisms function in nature. Note that spin estimates are only possible for the stellar mass black holes in X-ray binaries \citep{2021ARA&A..59..117R}. Existing samples strongly indicate that black holes of lower masses receive larger natal kicks \citep{Atri2019}, but the current sample size is small. If confirmed, this would support the notion that sources at the low mass end of the stellar mass black hole spectrum form via a temporary neutron star followed by fallback accretion, while the heavier black holes form via prompt collapse of a late-stage massive star \citep{1999ApJ...522..413F}.   

In order to provide meaningful constraints on their natal kicks, proper motions with uncertainties of order 1 masec/year are required. At that level of uncertainty, typical proper motions will be accurate to about the level of the local dispersion in stellar velocities (i.e. 40 km sec$^{-1}$) for a distance of 8 kpc.  To obtain 1 masec/year proper motion accuracy with 60 masec angular resolution requires a signal to noise of $60\sqrt{2}$ assuming a time baseline of 1 year. This is achievable for $I=23$ mag objects in about 25 minutes in each epoch with Lazuli (with fainter objects requiring some combination of longer time baselines or longer exposures). As most black hole X-ray binaries are too faint in quiescence, samples are currently severely limited, but Lazuli would allow at least tripling of current samples \citep{BlackCAT}. Importantly, the Lazuli field of view is sufficiently large that it should always contain bright quasars to use as background reference sources for astrometry; this is not in general the case for Hubble and Webb, which is a limiting factor in their ability to do precision astrometry.  

\begin{figure}
    \centering
    \includegraphics[width=1\linewidth]{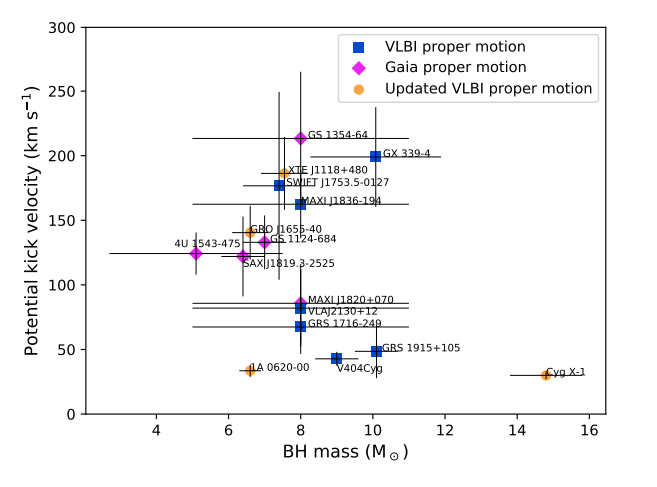}
    \caption{Potential  kick velocities versus black hole mass for the X-ray binaries for which both quantities are well-estimated. Figure from \citet{Atri2019}.}
    \label{fig:atri}
\end{figure}

\subsubsection{Globular cluster compact and ultracompact binaries}

Globular clusters (GCs), which can reach stellar densities as high as $10^6$ stars pc$^{-3}$, harbor a large amount of stellar remnants, many of which are in binaries with very short orbits, the so-called ``ultracompact binaries". These binaries can have a primordial origin, and some populations could be enhanced through dynamical mechanisms. Consequently, GCs are expected to be rich sources of gravitational wave emission \citep{2018Kremer}. For example, it is often suggested that a large fraction of the binary black holes expected to be found by LISA form in globular clusters \citep{2025Xuan}. A large number of short period accreting white dwarfs is also expected due to three body exchanges, tidal capture and direct collisions \citep{2006Ivanova}, and most of the known X-ray binaries with orbital periods less than 30 min are also in globular clusters. However, no short period double WD systems have been confirmed in globular clusters to date.

\begin{figure*}
    \centering \includegraphics[width=0.45\linewidth]{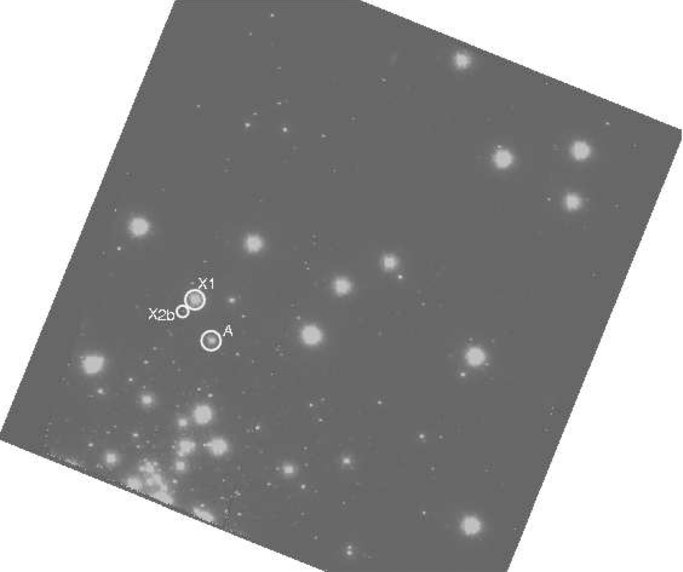}\includegraphics[width=0.35\linewidth]{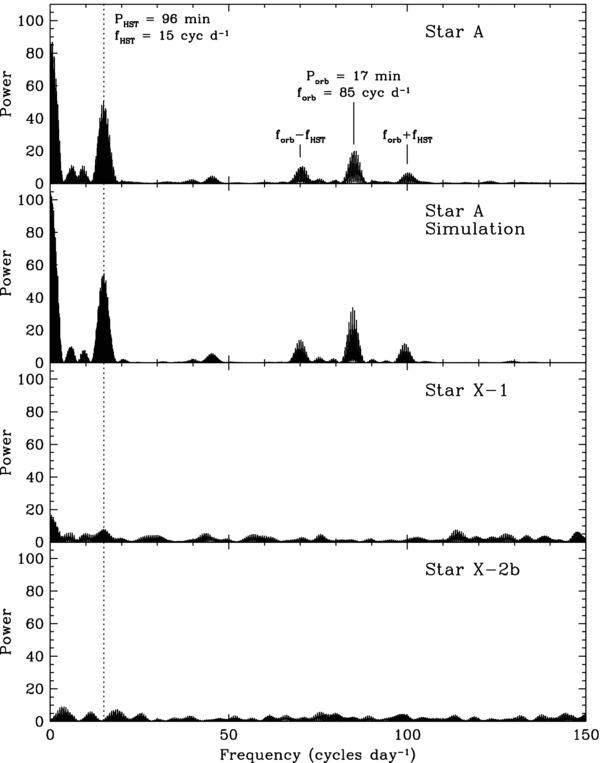}
    \caption{Left: A Hubble far-UV image of the globular cluster NGC~1851, illustrating that even in this very blue band, spatial resolution comparable to Hubble is needed to obtain good photometry of globular cluster objects.  Right: The power density spectrum of an ultracompact X-ray binary in NGC~1851, with a 17-minute period.  The aliasing illustrates the challenges of doing this work with a satellite in low Earth orbit, and was manageable only because the data were taken on three separate days.  Lazuli's ability to make long continuous observations will greatly simplify searches for short periods. Figure from \citet{Zurek2009}.}
    \label{fig:zurek}
\end{figure*}

Until now, the workhorse telescope to study compact binaries in GCs in the UV and optical band has been the Hubble Space Telescope due to its spatial resolution, which can resolve stars in the dense cores of GCs and help to identify periods and variable systems \citep{2021Belloni}. However, HST's orbital period of 90 minutes leads to severe aliasing in searches for short orbital periods, like the ones in ultracompact binaries -- see Figure \ref{fig:zurek}. It is furthermore inefficient at searching for rapid variability over a whole cluster, as none of its instruments combine fields of view greater than 1' with readout times less than 1 minute.
JWST faces a different set of problems, in that globular clusters are very crowded in red light, making it hard to detect faint blue variables.  

Lazuli's subarcsecond resolution and fast optical photometric capabilities will be able to determine orbital periods for compact binaries by performing extended and uninterrupted monitoring studies of GCs using the $u'$ and $g'$ bands to avoid contamination from the evolved and redder stars as much as possible. Lazuli data in narrowband filters such as H$\alpha$ and He\textsc{ii} will also allow the identification of systems through photometric excesses and deficiencies with respect to main sequence stars in the clusters \citep[e.g.][]{2017MNRAS.466..163W, 2018RS}. These observations can then be compared with Galactic disk populations, allowing a real understanding of the role of dynamical formation of close binaries in clusters, and opening up discovery space for LISA counterparts in environments with known distances. 

The determination of periodicities and excesses/deficiencies will allow detections of interacting compact binaries without having a prior detection in X-rays \citep[e.g.][]{2023PM}. This is an important factor to make better comparisons to population synthesis models, considering that most of the compact binary populations in GCs identified so far are X-ray biased, but most systems are expected to be faint and below the current detection limits of X-ray missions. Furthermore, non-X-ray biased variability studies would open up the opportunity to study GCs that so far have been poorly explored given their lack of X-ray sources. The rapid cadence of Lazuli will also permit to uncover the origin of many unclassified X-ray sources through period determination in some of the very well studied globular clusters with the Chandra observatory. 

\subsubsection{Field ultracompact binaries}

Ultracompact white dwarf (UCWD) binaries, i.e. systems where the primaries are WDs with orbital periods less than 1 hr are some of the populations that are expected to be detected both as individual systems and as part of the gravitational wave background by space interferometers like LISA \citep{2023Amaro-Seoane}. These systems, both with and without accretion, are faint and have proven difficult to identify. 

A significant fraction of the nearby accreting UCWD population has been identified through their transient behavior by optical all-sky surveys. Most of the short-period non-interacting systems have been identified through eclipses. Unfortunately current sky surveys are limited in depth to $\sim$21 mag, while most of the ultracompact white dwarf binaries are expected to be fainter than that, including some that would be easily detectable by LISA. 

Surveys such as Rubin's LSST are expected to unveil many systems, but their cadence is not optimal for period determination. Because sampling with 3-day average time resolution yields poor sensitivity to periods of sub-hour binaries, detached systems may show up as eclipsing objects on the white dwarf sequence long before their periods can be measured. In these cases, periods could be obtained quickly with Lazuli observations. As these systems can be very faint electromagnetically while being very strong gravitational wave sources, Lazuli has the opportunity to grow the sample of LISA verification binaries.  

As an example, the 6.91-minute period binary ZTF~1539+5027 would be detected with a wave amplitude about 100 times the noise limit of LISA. This source shows eclipses of nearly 100\% depth over a few percent of its orbit, and peaks at about 20th magnitude \citep{Burdge2019,LISA_astrophysics}. A similar system located 5 times further away would have a peak brightness of about 23.5 magnitude. It would still be detectable by LISA, but at a distance of 12.5 kpc rather than 2.5 kpc. If located in an unreddened part of the Galaxy, it  would show strong variability in LSST and be a good candidate for Lazuli follow-up, which will yield a SNR of 22 in the $g$-band in $20$ seconds of exposure time. This allows period determination and eclipse mapping to obtain radii and temperatures for both white dwarfs. 

Lazuli can be even more instrumental in understanding the broader population of double white dwarf binaries, detecting several hundred more systems than currently known within a few kpc, in particular when working synergistically with surveys like LSST. For example, LSST will detect mass-transferring UCWD binaries (e.g. AM CVns, He-CVs) with periods longer than about 20 minutes as transient sources (at shorter periods, systems are  persistently high state accretors). The outburst durations of strong candidate ultracompact sources, in contrast with other types of accreting white dwarfs like dwarf novae, can be identified by pre-outburst colors and brightnesses -- these objects should be consistent in color space with being reddened white dwarfs when in quiescence. The distinct recurrence times, rise and decline rates, and transient behavior duration \citep[e.g.][]{2026Kara} will allow discriminating UCWD binaries among the Rubin alert stream. 

\begin{figure}
    \centering
    \includegraphics[width=1\linewidth]{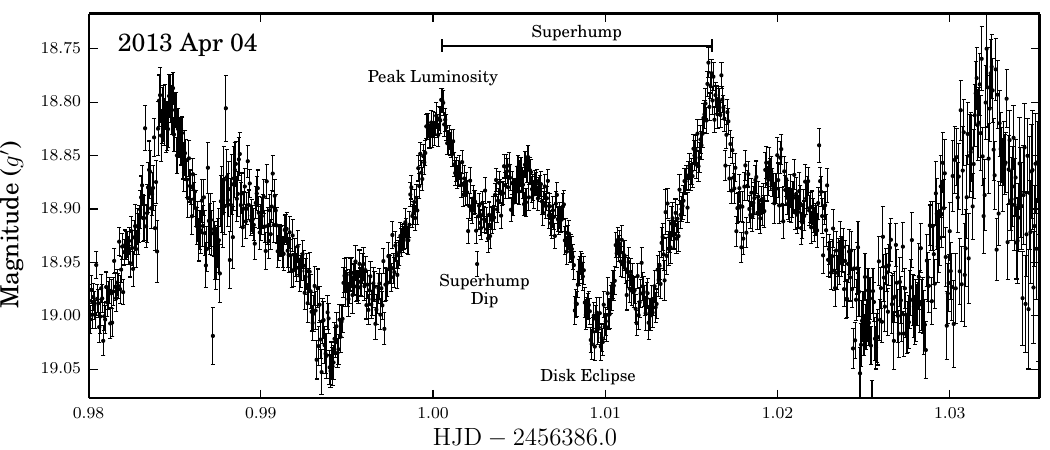}
    \caption{A rapid light curve of PTF~J191909.19+481506.2 from \citet{2014ApJ...785..114L}, illustrating superhumps that can be used to identify orbital periods in AM~CVn systems in outburst.}
    \label{fig:superhumps}
\end{figure}

However, the bulk of the candidates will need both rapid response and high cadence photometric follow-up to uncover their short periods during superoutbursts, when the systems are brightest. Superoutbursts are events longer and more energetic than "normal" outbursts and typically last $\approx$ 5 days \citep{TACOS}, meaning that a response of 1-2 days will be required after detection by LSST, given its typical three day cadence, or the source will have faded significantly and require much longer observations. With its fast slew capabilities and sensitivity, reaching $r' \approx 25.4$ mag in 60 seconds, Lazuli will be in a unique position to measure superhump periods. These are modulations (beat periods between the orbital period and much longer precessions in the outer disk) developed during superoutbursts that are generally within a few percent of the orbital period \citep{Smak2023} -- see Figure \ref{fig:superhumps} for an example -- and are then a good indicator for accretion and population synthesis studies.

Going from the sensitivity of surveys like ZTF and ASAS-SN to surveys like LSST should allow going from a few events per year at present, generally within 3~kpc, to dozens of events per year, out to $\sim$ 5 kpc, depending on reddening along a given line of sight. 
The amplitudes of variability are typically about 10\%, but because of stochastic flickering, several orbital cycles should be observed to ensure robust period determination. This requires observation durations of typically $\sim$5 hours, regardless of brightness. These observations, in turn, will allow the development of a real understanding of the orbital period distributions of AM CVn systems and a comparison of their properties with the predictions from purely gravitational-wave driven evolution, to determine (for example) whether other effects like magnetic braking \citep{2010arXiv1006.4112F} are an important ingredient of binary evolution.

\begin{figure*}
    \centering
    \includegraphics[width=1\linewidth]{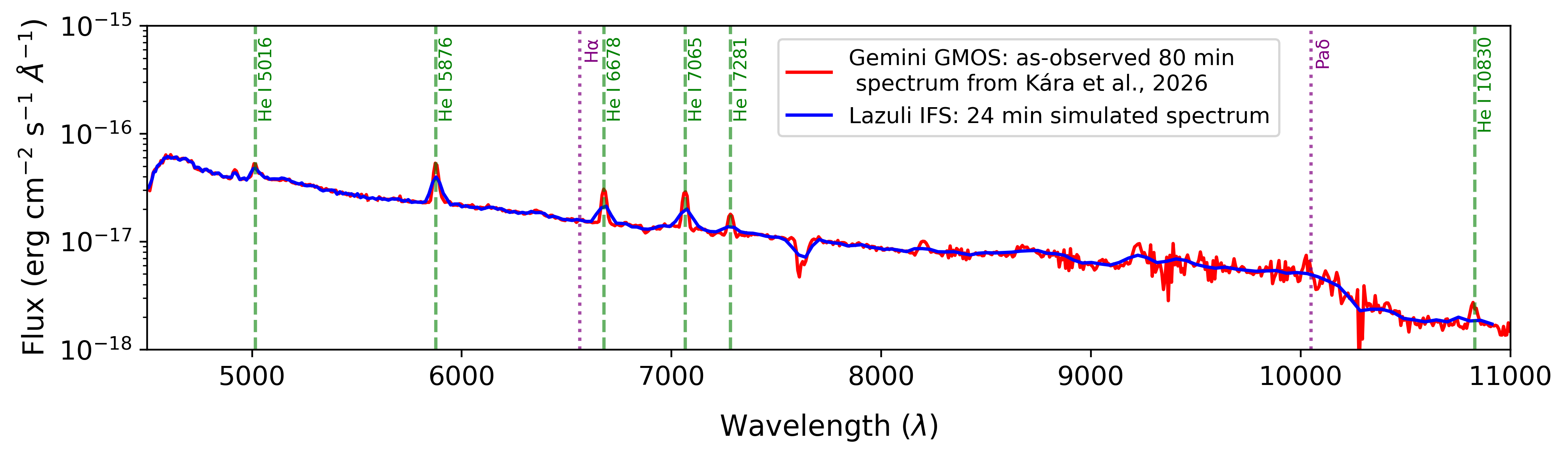}
    \caption{Spectra of the AM CVn ASASSN-19rg detected with a quiescence magnitude of G=20.4 and an orbital period of 44 min. The Gemini spectra required 80 min of GMOS \citep{2026Kara}, while Lazuli spectra required an exposure of 24 min. The characteristic He emission lines are well detected even when the exposure time is substantially shorter.}
    \label{fig:spectra_lazuli}
\end{figure*}

The spectra of ultracompact binaries will be essential to fully confirm their nature and determine their properties, in particular for very faint systems detected by surveys like LSST. In this context Lazuli can revolutionize the field, as spectra can be obtained with substantially shorter exposure times compared to existing large ground based facilities (see Figure \ref{fig:spectra_lazuli}). This will provide constraints on temperatures from the continuum shape and the presence of accretion-related emission lines. Also, Lazuli's broader wavelength coverage compared to existing facilities (e.g. Gemini-GMOS) will enable studies of the disk and donor contributions using the redder ends of the spectra. 

At the present time, there is a modest-sized, but rapidly growing sample of double-white dwarf binaries which have predicted gravitational wave strains that should make them detectable with LISA.  What has not been established is the extent to which the evolution of these objects is driven entirely by gravitational radiation. They can be affected by tides (e.g. \citealt{2025arXiv250721821M, 2012ApJ...745..137V}, magnetic braking \citep{2010arXiv1006.4112F, 2024MNRAS.529L..28M}, and tertiary companions (something which has been suggested for an ultracompact X-ray binary which is a LISA test binary \citealt{2001ApJ...563..934C}). A greater understanding of the period derivatives of these fast-orbiting systems is needed to ensure that good template banks are produced in advance of LISA.  

For one of the most extreme double white dwarfs known, a tidally-induced component of the period derivative of about $3.5\times10^{-12}$ sec/sec is expected.  This component would be expected to produce a timing offset relative to the predictions of pure gravitational radiation of about 0.15 milliseconds on a 3 year timeline, so over a potential 10 year mission, they could be expected to produce residuals of order half a millisecond. In some cases there is clear evidence that magnetic torques are also at play \citep{2005MNRAS.363..581M}, with much larger period derivatives seen due to this effect.  Both effects would be expected to be strongest in the closest binaries, and continuing time domain surveys can be expected to find binaries with even shorter periods than currently known. Work on the faintest members of the class, which represent the majority expected, requires sensitive telescopes capable of performing fast readouts and long, uninterrupted observations.

LISA is also expected to detect double white dwarf binaries in the Magellanic Clouds and dwarf spheroidal Milky Way neighbors \citep{Korol_dwarfs, LMC_LISA}.  In many cases, these objects will have faint optical counterparts, and this is likely to be most severe for the binaries with the heaviest white dwarfs, which would potentially be supra-Chandrasekhar mergers. If the Lazuli mission lasts long enough to overlap with LISA, then ultra-deep follow-up of LISA binaries could be done, but given the expectation of about 100 objects in the LMC \citep{LMC_LISA}, searches could be made in the regions with the highest stellar densities, likely revealing a few eclipsing objects ahead of time.  Typical LMC white dwarfs will be about $28^{th}$ magnitude, so deep data will be needed; multiple sensors can be used in parallel to cover large solid angle, as the fainter double white dwarfs should have temperatures such that they will be well-detected over most optical bands.

\subsubsection{Field binary stars as probes of dark matter}

At the present time, most approaches to understanding the distribution of dark matter in galaxies focus on measures of the gravitational potential as a function of position, either through virial measurements in elliptical galaxies or rotation curves in disk galaxies.   The local gravitational potential in the Milky Way can now be measured using the timing of binary systems, either via pulsar timing \citep{Donlon2024} or eclipse timing \citep{Chakrabarti2022}.  

The core signal in both cases is the period derivative of the binary being observed, which can be affected either by {\it bona fide} changes in its period, or by the Doppler effect due to acceleration in the Galactic potential, as well as a few other effects for which one can correct in a straightforward manner.  The binaries used for this work must be sufficiently wide that tidal effects within the binary do not affect their orbital evolution.  This thus places a constraint that the binaries be relatively long period, significantly longer than the satellite orbit for HST.  Then, because the centroiding precision scales linearly with the eclipse duration and inversely with the signal-to-noise \citep{2010exop.book...55W}, high precision is needed.

The ideal sample for this work are the $\sim$ 200 wide binaries already well characterized by Kepler, for which only a single additional eclipse is needed if it is very well measured \citep{Chakrabarti2022}, and typical eclipse durations are a few hours each \citep{Chakrabarti2022}, with periods typically of tens of days.  Measurements of the Galactic potential sufficient to test whether the Milky Way is in a static potential and to constrain effects like clumping of dark matter require timing of eclipses to a precision of about 0.1 seconds \citep{2019PASA...36...38S,Chakrabarti2020,Chakrabarti2022}.

\citet{Chakrabarti2022} give as an example of these systems the 23-day period eclipsing binary KIC 4144236.  For this system, a 5-hour Lazuli observation taken around the eclipse would have 300 1-minute observations.  Given its brightness of $V=11.6$, each observation in $g$ would have signal-to-noise of about 8700, with Lazuli.  This, in turn, would yield a centroid of the eclipse time a bit better than 0.1 seconds with a single eclipse, which gives a precision on the acceleration of about $10^{-7}$ cm/s$^2$, which is of the same order as the variance seen from the accelerations of a small number of local pulsars \citep{2021ApJ...907L..26C}; data on additional eclipses can yield precision that will scale with the square root of the number of measurements.  Roughly similar precision can be obtained with HST, but this imposes challenging scheduling considerations to ensure that the Hubble visibility windows line up with the times of the eclipses.  Lazuli's large field of regard and continuous visibility makes the scheduling substantially more efficient, especially to observe a sample of objects in the Kepler field at different distances from the Sun.

\section{Synergies and coordination with other facilities}
\label{sec:synergy}
The Lazuli space observatory will start operations at an opportune time to enable key capabilities for time-domain and multi-messenger astronomy, in a landscape of observatories that will provide an unprecedented volume of new discoveries in the transient sky, including Rubin, Argus, DSA, Roman, ULTRASAT, UVEX, and Einstein Probe. 

Time-domain science inherently depends on coordinated workflows -- from discovery and classification to rapid spectroscopic response and long baseline monitoring. Lazuli is being designed to contribute at the critical stage where early characterization most directly constrains progenitor physics and energy sources.
The combination of rapid response with a large collecting area opens up a part of parameter space that has to date only been probed superficially by 8m class ground-based telescopes. However, these facilities suffer from weather and technical downtime, as well as a low duty cycle (operating only during astronomical night time). While small telescope networks can provide near-continuous coverage, their sensitivity is typically limited to very nearby events. 

Lazuli has the potential to open up this parameter space by providing a low latency response with flagship-class sensitivity. Moreover, the dynamic scheduling system makes it possible to adjust observing strategies in near-real-time. Its lunar synchronous orbit with very limited obstructions due to the Moon and Earth systems means that it is capable of very high cadence repeat visits for extended periods of time, anywhere in its field of regard. 

With multiple next-generation surveys ramping up in the second half of the decade, the next few years represent a unique opportunity to establish coordinated infrastructure that can fully exploit their combined capabilities. 

\subsection{The short timescale frontier}
Argus will provide transient alerts from higher cadence observations (down to $\sim$ minute timescales) than any other transient survey to date, over a very large instantaneous field of view of 8000 sq. deg \citep{2022PASP..134c5003L}. Similarly, ULTRASAT will perform high cadence, wide-field surveys (200 sq. deg. field of view) at UV wavelengths. With these facilities pushing the discovery phase space to unprecedented timescales over large volumes, Lazuli's rapid-response optical/IR observations (on target in $\sim$hours following a ToO trigger) will enable the community to fully open up this under-explored parameter space with broad-band imaging and low-resolution spectroscopic follow-up. 
These synergies extend naturally into the multi-messenger domain: the same rapid-response architecture that enables early optical/IR spectra will also facilitate prompt follow-up of gravitational-wave triggers, high-energy transients such as the rapidly growing sample of fast X-ray transients detected by Einstein Probe \citep{2022hxga.book...86Y} and radio observatories (e.g. the Deep Synoptic Array, \citealt{2019BAAS...51g.255H}). The next observing run of the LIGO-Virgo-Kagra GW detector network is currently scheduled for 2028 and expected to last several years, well aligned with the primary mission lifetime of Lazuli.

\subsection{The high redshift frontier}
In addition to the rapid response mode, Lazuli's spectroscopic sensitivity out to 1.7$\mu$m and broad-band imaging makes it a powerful tool to characterize transient discoveries at higher redshifts. 
It will be capable of follow-up observations over an instantaneous volume that encompasses the discovery volume of Rubin's LSST single-visit depth ($g \approx 25$ mag, \citealt{2022ApJS..258....1B}) with both photometry (high SNR expected in $\sim$ minute exposures) and spectroscopy ($\approx$ 1 hour exposure time for SNR$\approx10$ at $g' = 24$ mag). Similarly, the Roman space telescope high latitude wide area survey single dither (pass) depths\footnote{https://roman-docs.stsci.edu/roman-community-defined-surveys/high-latitude-wide-area-survey} of $\sim$25.5 (26) mag in the IR bands are well-matched to spectroscopic follow up of high-z transients with Lazuli in exposure times of $\approx$hours. 

Its sensitivity will further enable time-series measurements of sources down to 28 mag in exposure times of 3-4 hours, which opens up the opportunity to photometrically characterize samples of newly discovered transients at very high redshift. This is a frontier that is being opened with very deep but small area imaging programs on JWST (e.g. \citealt{2025ApJ...979..250D, 2026arXiv260108931F}), and will be more accessible in the near future through Rubin and Roman deep (survey) fields. This will enable the classification and characterization of transient events with low volumetric rates, including lensed supernovae, pair instability and electron-capture supernovae, as well as spectroscopic follow-up of a sample of $z > 1$ supernovae to constrain the evolution of SN rates and properties over cosmic time (e.g. \citealt{2025ApJ...981L...9P}). \\

Ultimately, the greatest scientific gains in TDAMM astronomy will come not from individual observatories operating in isolation, but from a coordinated network of facilities that can discover, characterize, and contextualize transient events in near real time.
In combination with other facilities, the community will have at its disposal a fleet of both small (SVOM, Einstein Probe, COSI) and large space telescopes (including Lazuli) capable of rapid turn-around follow-up observations from the X-ray through NIR bands. 

\section{Summary}
\label{sec:summary}
The fields of time domain astronomy, variability studies, and multi-messenger science are entering a discovery rich era where a main bottleneck will be the ability to obtain detailed follow-up observations. A key limitation in this landscape is the inability to rapidly respond to faint, fast-evolving transients with multi-band imaging and spectroscopy afforded by a large aperture space telescope. 

The Lazuli space observatory will combine a large collecting area with large-format CMOS imaging sensors and low-resolution integral field spectroscopy, a rapid response capability (4 hr requirement, 90 min goal) and flexible scheduling. This opens up parts of parameter space that are currently very difficult to access, including early-time spectroscopy of transients, sensitive high-frequency imaging, long (12--24 hr) uninterrupted observations at high (seconds to minutes) cadence, accurate astrometry, and very deep field imaging. 

Lazuli's capabilities translate into science opportunities for a wide range of physical phenomena. These include the very early-time follow-up of multi-messenger sources and young, explosive and fast transients to constrain the progenitors, early-time physics, heavy element nucleosynthesis, central engines, and explosion mechanisms of a wide variety of extragalactic sources; constraints on cosmological themes including the Hubble constant through multi-messenger standard sirens and independent calibrations through strongly lensed supernovae; measuring time-lags in accretion binaries to constrain jet physics and eclipse timing of wide binaries to constrain the local dark matter distribution; constraining natal kicks of BH and NS binaries through proper motions and astrometry; and discovery of field and globular cluster binaries through deep, time-resolved imaging. 

As part of the Schmidt Observatory System, Lazuli's capabilities will enable strong synergies with the short timescales to which Argus and DSA are uniquely sensitive, both through coordinated observations (e.g. contemporaneous observations of repeating FRBs with DSA and Lazuli) as well as by providing the community with an integrated framework for early discovery and rapid response, large aperture follow-up from space.

\begin{acknowledgments}
We thank M. Bulla, J. H. Gillanders, P. Nugent, A. Rest, S. Suyu, and S. Taubenberger for discussions.  DJS acknowledges E. Beasor and J. Jencson for providing the outburst spectrum simulated in Figure~\ref{fig:RSGoutburst}; DJS also acknowledges Y. Dong for discussions on precursor outbursts. 
The Lazuli Space Observatory is part of the Eric \& Wendy Schmidt Observatory System, an initiative of Schmidt Sciences.
\end{acknowledgments}

\bibliography{bibliography}{}
\bibliographystyle{aasjournalv7}

\end{document}